\documentclass[iop]{emulateapj}

\usepackage{graphicx}
\usepackage{amsmath,amssymb}
\usepackage{url}
\newcommand{\xmm}{{\it XMM}}
\newcommand{\swift}{{\it Swift}}
\newcommand{\chandra}{{\it Chandra}}
\newcommand{\xmmn}{{\it XMM-Newton}}
\newcommand{\Msun}{$M_{\odot}$}
\def\ergs{~{\rm erg~s^{-1}}}
\def\deg{\ensuremath{^{\circ}}}

\shorttitle{Crossing the Eddington limit}
\shortauthors{A. D. Sutton et al.}

\begin{document}

\title{Crossing the Eddington limit: examining disk spectra at high accretion rates}
\author{Andrew D Sutton\altaffilmark{1}}
\author{Douglas A Swartz\altaffilmark{1}}
\author{Timothy P Roberts\altaffilmark{2}}
\author{Matthew J Middleton\altaffilmark{3}}
\author{Roberto Soria\altaffilmark{4,5}}
\author{Chris Done\altaffilmark{2}}

\affil{\altaffilmark{1}Astrophysics Office, NASA Marshall Space Flight Center, ZP12, Huntsville, Al 35812, USA}
\affil{\altaffilmark{2}Centre for Extragalactic Astronomy, Department of Physics, Durham University, South Road, Durham, DH1 3LE, UK}
\affil{\altaffilmark{3}Institute of Astronomy, University of Cambridge, Madingley Road, Cambridge, CB3 0HA, UK}
\affil{\altaffilmark{4}International Centre for Radio Astronomy Research, Curtin University, GPO Box U1987, Perth, WA 6845, Australia}
\affil{\altaffilmark{5}Sydney Institute for Astronomy, School of Physics A28, The University of Sydney, Sydney, NSW 2006, Australia}
\email{andrew.d.sutton@nasa.gov}

\begin{abstract}

The faintest ultraluminous X-ray sources (ULXs), those with 0.3--10 keV luminosities 
$1< L_{\rm X}/10^{39} <3~\ergs$, tend to have X-ray spectra that are disk-like but 
broader than expected for thin accretion disks. These `broadened disk' spectra are 
thought to indicate near- or mildly super-Eddington accretion onto stellar remnant 
black holes. Here we report that a sample of bright thermal-dominant black hole 
binaries, which have Eddington ratios constrained to moderate values, also show broadened 
disk spectra in the 0.3--10 keV band at an order of magnitude lower luminosities. 
This broadening would be missed in studies that only look above $\sim 2~{\rm keV}$. 
While this may suggest that broadened disk ULXs could be powered by accretion onto 
massive stellar remnant black holes with close to maximal spin, we argue in favor of 
a scenario where they are at close to the Eddington luminosity, such that radiation 
pressure would be expected to result in geometrically slim, advective accretion disks. 
However, this implies that an additional physical mechanism is required to produce 
the observed broad spectra at low Eddington ratios.

\end{abstract}

\section{Introduction}

Ultraluminous X-ray sources (ULXs) are non-nuclear point sources in external galaxies that are primarily defined 
by their extraordinarily high X-ray luminosities of $L_{\rm X} \ge 10^{39} \ergs$. To produce such extreme luminosities, most ULXs 
are presumably powered by accretion onto black holes (although as of writing three have been reported to contain neutron stars, 
\citealt{bachetti_etal_2014,  fuerst_etal_2016, israel_etal_2016a, israel_etal_2016b}). This lower 
limit in luminosity is close to the Eddington 
luminosity, $L_{\rm Edd} \sim 1.3 \times 10^{39} \ergs$,  of a typical $\sim 10$~{\Msun} stellar mass black hole. 
Thus ULXs must contain larger 
black holes and/or have higher Eddington ratios, $l \equiv L/L_{\rm Edd}$, than we commonly see in Galactic black hole binaries (BHBs).

There are several well-defined accretion states of Galactic BHBs 
that tend to occur within certain ranges of (sub-)Eddington ratios. 
For instance, quiescence and the hard state have similar properties 
(a hard power-law spectrum with a cut-off at a few tens of keV) and occur at $l \ll 1$ (although the hard state 
has been reported to peak at high fractions of the Eddington luminosity, $l \lesssim 0.7$, during a few outbursts, \citealt{dunn_etal_2010}). 
On the other hand, the thermal dominant (TD) state typically occurs at higher Eddington ratios, $0.1 \lesssim l \lesssim 0.3$. 
BHBs are also observed in a less well understood steep power-law state, which is characterized by a power-law spectrum with $\Gamma > 2.4$,
and occurs at both higher and lower luminosity than the TD state 
(for more in-depth reviews of sub-Eddington accretion states see e.g., \citealt{mcclintock_and_remillard_2006,remillard_and_mcclintock_2006,done_etal_2007}). 
Thus, it was initially hoped that rudimentary constraints could be placed on Eddington ratios, hence
black hole masses, in ULXs based on their energy spectra alone, if they were in standard sub-Eddington states.

However, most of the highest quality 
ULX spectra 
show near ubiquitous high energy curvature at around 5 keV, which is often coupled to a soft excess 
\citep{stobbart_etal_2006,gladstone_etal_2009,bachetti_etal_2013,walton_etal_2014}. 
We note 
that a TD spectrum could appear as an absorbed cut-off power-law, without a soft excess, in the $\sim 0.3$--10~keV band. 
However, TD-like spectra can be rejected for many ULXs, and the high black hole masses implied by sub-Eddington states 
are often inconsistent with the $\ge 1~{\rm keV}$ disk temperatures obtained when ULX spectra are fitted with disk models (e.g., \citealt{gladstone_etal_2009}). 
Exceptions to this may include the least luminous ULXs (e.g., a 30~$M_{\odot}$ black hole in the TD state at $0.3 L_{\rm Edd}$ would have a bolometric luminosity of 
$\sim 1.2 \times 10^{39}~{\rm erg~s^{-1}}$ and a disk spectrum peaking at $\sim 1~{\rm keV}$; \citealt{shakura_and_sunyaev_1973}).

This combination of a soft excess and high energy curvature, along with the high luminosities, is unlike any known sub-Eddington state, 
leading to the suggestion of a 
new accretion state, which is generally termed the `ultraluminous state' \citep{gladstone_etal_2009}. 
The ultraluminous state has been further subdivided into three distinctive spectral 
types, which we now refer to as the broadened disk (BD), hard ultraluminous 
and soft ultraluminous regimes \citep{sutton_etal_2013b}.
The luminosity of the majority of BD ULXs ($\lesssim 3 \times 10^{39} \ergs$) is such that they bridge the gap 
between the sub-Eddington BHBs and the hard and soft ultraluminous sources, which has led to 
the suggestion that they may represent accretion at $l \sim 1$ 
while 
ULXs in the hard and soft ultraluminous regimes,
seen 
almost exclusively at luminosities above $\sim 3 \times 10^{39} \ergs$, likely represent 
super-Eddington mass accretion rates \citep{sutton_etal_2013b}. 
Hard and soft ultraluminous sources have two distinct spectral components 
that may indicate the presence of radiatively-driven outflows in the supercritical accretion regime; 
the disk-like spectra in 
BD sources could perhaps be intermediate in shape between these two-component spectra and a disk-only spectrum \citep{middleton_etal_2011b,middleton_etal_2012}.

Sub-Eddington TD spectra are dominated by emission from a geometrically thin accretion disk, which can be approximated by a 
multi-color disk 
(MCD) model (plus a faint high energy power-law tail; e.g., \citealt{mcclintock_and_remillard_2006}). 
As $l \rightarrow 1$, radiation pressure would be expected to cause 
the disk scale height to increase, thus advection should start to become important (e.g., \citealt{abramowicz_etal_1988,watarai_etal_2001,sadowski_2009}) 
giving rise to the so-called slim disks with spectra that are subtly broader than the TD state; perhaps consistent with the BD ULX state 
(a steep power-law  state has been reported between the thin and slim disk regimes in XTE 1550$-$564, but this is short-lived; \citealt{kubota_and_makishima_2004}). 
Recently it has been noted that some $l<1$ TD BHBs have spectra that are subtly broader than even the best current thin disk models. 
For example, {\sc bhspec} \citep{davis_etal_2005} includes both the changing color temperature correction due to atomic features and 
relativistic effects, and is intrinsically broader than {\sc kerrbb} (relativistic effects, but constant color temperature correction) or an MCD (no relativistic effects and a constant color temperature correction, see \citealt{done_and_davis_2008}). 
Nonetheless, 
{\it Suzaku} spectra of LMC X-3 \citep{kubota_etal_2010} and
{\xmmn} spectra of GX~339$-$4 
\citep{kolehmainen_etal_2011} are broader than can be reproduced by the {\sc bhspec} disk model. 
\cite{kolehmainen_etal_2011} show that the spectra of GX~339$-$4  
are best fitted by two-component phenomenological models 
that are strikingly similar to those used to argue in favor of two-component
spectra in many ULXs; namely, a low temperature MCD plus cool, optically thick Comptonization, e.g., \cite{gladstone_etal_2009}. 
Similarly, unusual broadening is seen in a transient ULX in M31 (\citealt{straub_etal_2013}; M31 ULX1 in this study). 
This source has a spectrum that 
is well-fitted by slim disk models when it is at 
ULX luminosities and in the BD regime. 
Interestingly, as this source faded to lower luminosities, its spectrum remained unusually broad, such that standard thin disk 
models were unable to provide a good fit to the data.

Further similarities between the TD BHBs and BD ULXs are evident from fitting irradiated disk models to X-ray and optical data.
In BHBs, X-rays from the inner regions of the accretion disk irradiate the outer disk, where they are reprocessed and re-emitted 
as optical/UV emission at the local black body temperature \citep{gierlinski_etal_2008, gierlinski_etal_2009}. As thin 
disks transit to slim disks with increasing Eddington ratio, the fraction of emission that is reprocessed would be expected to decrease, due 
to self-shielding in the accretion disk \citep{kaaret_and_corbel_2009}. However, it has been reported that BD ULXs have similar reprocessing 
fractions to the TD BHBs, suggesting their geometries may not be vastly different \citep{sutton_etal_2014}.

Clearly something is missing from our understanding of 
accretion disks. 
Motivated by the 
possible similarities between sub-Eddington TD BHBs and potentially $\sim$ Eddington BD ULXs,
we here present a comparison of TD and BD sources. 
Historically, Galactic BHBs and ULXs have mainly been studied with different instruments (e.g {\it RXTE} and 
{\it XMM-Newton}/{\it Chandra}, respectively), which crucially have different bandpasses. 
Although external galaxies have been observed in the $\sim 0.3$--10 keV range (e.g. \citealt{vulic_etal_2016}), 
the net counts obtained from binaries are generally not sufficient to allow spectral studies comparable to those 
that are possible in Galactic BHBs, or even nearby ULXs. There are also additional complications coming from the difficulties in distinguishing 
BHBs from neutron star binaries. 
This bandpass mis-match presents serious, but 
often overlooked, limitations in comparisons 
between the two source classes, as subtle broadening may not be evident if disk spectra 
are only observed above 2~keV. 
As such, we here assess high count rate $\sim 0.3$--10~keV CCD spectra of nearby BHBs (of known black hole mass) 
to establish whether they can truly be described by TD models while at substantially sub-Eddington luminosities 
or if an additional component, especially broadening, is required. 
Comparison is also made to more luminous (but poorer spectral quality) ULXs in nearby galaxies 
to better understand the accretion physics and Eddington ratio giving rise to the BD ULX state.
We describe the selection of our TD and BD samples in Section \ref{sampleselect}, 
outline the tests we carry out in Section \ref{analysis}, present our findings in Section \ref{results} and discuss the implications of these 
in the context of the broader literature in Section \ref{discussion}.

\section{Sample selection}\label{sampleselect}

The intention of this work was to compare BD ULXs with high mass accretion rate BHBs. As such, we considered CCD observations 
in the $\sim$ 0.3--10 keV energy range of BHBs with 
$l \gtrsim 0.1$. 
Although BHBs are occasionally 
detected in a bright `steep power-law' state (e.g., \citealt{remillard_and_mcclintock_2006}), this state is characterized 
by a power-law-like spectrum, while the TD state appears most comparable to the BD ULXs. 
We therefore concentrated on the TD BHBs in this work.

We defined a sample of BHB observations taken from 
the 21 BHBs with known masses out to the distance of the LMC from \cite{zhang_2013}. 
This contains the same 20 confirmed BHBs as \cite{remillard_and_mcclintock_2006}, with the 
addition of H1743$-$322.
We obtained mission-long one day average {\it Rossi X-ray Timing Explorer} 
({\it RXTE}) All-Sky Monitor (ASM) light curves from the ASM 
web page\footnote{\url{http://xte.mit.edu/ASM\_lc.html}} for all of the sources. 
The maximum {\it RXTE} ASM count rate for each source was used to extrapolate an estimate of the peak 
Eddington ratio ($l_{0.3 \mbox{--} 10~{\rm keV}} \approx L_{0.3 \mbox{--} 10~{\rm keV}}/(M_{\rm BH} \times 1.3 \times 10^{38}~{\rm erg~s^{-1}}~M_{\odot}^{-1})$) using 
{\sc pimms},\footnote{\url{https://heasarc.gsfc.nasa.gov/\mbox{cgi-bin}/Tools/w3pimms/w3pimms.pl}} 
assuming a 1 keV black body shaped spectrum. 
We emphasize that this model is just a rough approximation of a thermal spectrum which we used to define our sample, and 
we would not expect it to be appropriate for accurately extrapolating {\it RXTE} fluxes to the 0.3--10 keV band, 
especially in spectra with strong power-law components.
We also note that we here define the Eddington ratio in terms of the 0.3--10 keV luminosity. 
While the bolometric correction for an absorbed 1~keV black body observed in the 0.3--10 keV range is 
only $\sim 1 \mbox{\%}$, larger corrections would be expected for more realistic spectra (e.g., \citealt{migliari_and_fender_2006} 
reported $L_{2\mbox{--}10~{\rm keV}}/L_{\rm Bol} \sim 0.8$ for outbursting BHBs, albeit in a narrower energy band). 
We did not correct luminosities for inclination in an accretion disk. 
This is motivated by the lack of good inclination estimates in many sources, especially the ULXs. 
However, we do note the approximate correction factors for the BHBs in the final sample (Table \ref{BHB_catalogue}), and 
for most estimates of their inclinations this is only a factor of 0.5--1.5.

For BHBs that exceeded 
$l \approx 0.1$ in the ASM light curves, 
we searched the High Energy Astrophysics Science Archive Research Center (HEASARC) 
archive\footnote{\url{https://heasarc.gsfc.nasa.gov/}} for {\swift} and {\xmmn} 
detections. Five sources had {\xmmn} and/or {\swift} detections during time periods 
where the ASM light curves indicated they met our Eddington ratio selection criteria. 
These were GRO 1655$-$40, GX~339$-$4, GRS 1915$+$104, LMC X-3, and LMC X-1.

\begin{table*}
\centering
\caption{The BHB sample}
\label{BHB_catalogue}
\begin{tabular}{cccccccccc}
\hline
Src ID$^a$ & ${M_{\rm BH}}^{b}$ & Reference & Distance$^c$ & Reference & ${N_{\rm H}}^d$   & $l^e$ & $i^f$           & Reference & ${1/(2\cos{i})}^g$ \\
           & $({M_\odot})$     &           & (kpc)        &           &$(10^{20}~{\rm cm^{-2}})$& & (deg.)       &           & \\
\hline
GX~339$-$4 & ${7.5 \pm 0.8}$   & 1         & $\sim 8$     & 4         & \ldots            & $0.2$ & 20--30, ${>45}$ & 5, 6 & 0.5--0.6, $>0.7$ \\
LMC X-3    & $6.98 \pm 0.56$   & 2         & $48 \pm 2$   & 3         & 4.68              & $0.9$ & ${\sim 70}$     & 7    & $\sim 1.5$ \\
LMC X-1    & $10.91 \pm 1.41$  & 3         & $48 \pm 2$   & 3         & 6.73              & 0.7   & ${\sim 36}$     & 3    & $\sim 0.6$ \\
\hline \\
\end{tabular}
\begin{minipage}{\linewidth}
Notes:
$^a$common source name;
$^b$black hole mass; 
$^c$distance to the BHB; 
$^d$Galactic neutral hydrogen column density in the direction of the LMC sources 
from \citealt{dickey_and_lockman_1990}, no 
value is given for GX~339$-$4 as it is within our Galaxy; 
$^e$approximate Eddington ratio ($L_{\rm 0.3\mbox{--}10~keV}/(M_{\rm BH} \times 1.3 \times 10^{38}~{\rm erg~s^{-1}}~M_{\odot}^{-1})$) 
at the peak {\it RXTE} ASM count rate for an 
assumed absorbed 1 keV blackbody shaped spectrum; 
$^f$inclination of the binary system; 
$^g$approximate correction factor between apparent luminosity and the luminosity of an accretion disk.
References: 
(1) \cite{chen_2011} (note that this mass estimate is based on a scaling of the spectral and timing properties, but it is in 
agreement with the mass function derived from the radial velocity curve,  $f(M_{\rm BH}) = 5.8 \pm 0.5~M_{\odot}$, \citealt{hynes_etal_2003});
(2) \cite{orosz_etal_2014};
(3) \cite{orosz_etal_2009}; 
(4) \cite{zdziarski_etal_2004}; 
(5) \cite{miller_etal_2004}; 
(6) \cite{kolehmainen_and_done_2010}; 
(7) \cite{orosz_etal_2014}.
\end{minipage}
\end{table*}

\begin{table*}
\centering
\caption{BHB observation log}
\label{BHB_obs}
\begin{tabular}{ccccccc}
\hline
Instrument/mode$^a$ & Obs ID$^b$ & Date$^c$ & ${\rm{t_{exp}}}^d$ & Count rate$^e$ & ${\rm log_{10}}(f_{\rm X})^{f}$ & $l^g$ \\
                    &            &          & (ks)               & $(\rm s^{-1})$ & $({\rm log_{10}(erg~cm^{-2}~s^{-1})})$ &       \\
\hline
\multicolumn{7}{c}{GX~339$-$4}\\
{\xmm}-B & 0093562701 ({\xmm}-1) & 2002 Aug 24 & 1.3 & $5.2 \times 10^3$ & $-7.548 \pm 0.002$ & 0.22 \\
{\xmm}-B & 0410581201 ({\xmm}-2) & 2007 Feb 19 & 0.4 & $5.2 \times 10^3$ & $-7.554 \pm 0.003$ & 0.22 \\
\hline
\multicolumn{7}{c}{LMC X-3}\\
{\xmm}-T & 0109090101 ({\xmm}-1) & 2000 Nov 24 & 9.2 & 420 & $-8.784 \pm 0.002$ & 0.46 \\
{\swift}-WT & 00037080006 ({\swift}-1) & 2007 Dec 02 & 9.2 & 25 & $-9.037 \pm 0.001$ & 0.25\\
{\xmm}-T & 0671420301 ({\xmm}-2) & 2011 May 27 & 8.0 & 250 & $-9.028 \pm 0.004$ & 0.26 \\
\hline
\multicolumn{7}{c}{LMC X-1}\\
{\xmm}-T & 0112900101 ({\xmm}-1) & 2000 Oct 21  & 4.9 & 120 & $-9.07 \pm 0.01$ & 0.16 \\
{\swift}-WT & 00037079002 ({\swift}-1) & 2007 Dec 06 & 9.8 & 14 & $-9.05 \pm 0.03$ & 0.17 \\
{\swift}-WT & 00037079003 ({\swift}-2) & 2007 Dec 10 & 4.4 & 15 & $-9.09 \pm 0.03$ & 0.16 \\
\hline \\
\end{tabular}
\begin{minipage}{\linewidth}
Notes:
$^a$instrument with which the observations were taken - {\xmmn}~EPIC in timing ({\xmm}-T) or burst ({\xmm}-B) mode, or {\swift} XRT in window timing mode ({\swift}-WT);
$^b${\xmmn} or {\swift} observation identification number;
$^c$observation start date;
$^d$effective good exposure time; 
$^e$net source count rate (to two significant figures); 
$^f$logarithm of the unabsorbed 0.3--10 keV flux 
averaged over the observation calculated from the {\sc xspec} model 
averaged over the observation calculated from the {\sc xspec} model 
$\textsc{tbabs} \times \textsc{( diskbb + comptt )}$ using the {\sc cflux} convolution model;
$^g$ approximate Eddington ratio ($l_{0.3 \mbox{--} 10~{\rm keV}}$).
\end{minipage}
\end{table*}

We excluded GRS 1915$+$104 from our sample, as its strong absorption column 
($N_{\rm H} \sim 4 \times 10^{22}~{\rm cm^{-2}}$, 
\citealt{ebisawa_1998}) makes detailed study of this source difficult below 2 keV. 
Additionally, GRO 1655$-$40 had a single {\it Swift} XRT observation in low rate photodiode mode 
that met our Eddington ratio criteria, but it also had a strong power-law component in its spectrum, 
even below 10 keV. This was also the case in {\xmmn} observations 0410581301 and 0023940401 of 
GX~339$-$4 and LMC X-1 respectively. As these three observations deviated from disk dominated spectra, we do not 
consider them further. 
Details of the remaining BHB sample are 
given in Table \ref{BHB_catalogue},\footnote{A distance of $\sim 50~{\rm kpc}$ to the LMC is currently preferred 
\citep{pietrztnski_etal_2013, degrijs_etal_2014, crandall_and_ratra_2015}, but we use 48 kpc here for consistency with 
the \cite{orosz_etal_2009} mass estimate of LMC X-1.} and an observation log in Table \ref{BHB_obs}. Although they were not initially 
flagged as such, it later became apparent that LMC X-1 {\swift}-1 and -2 may also have deviated from a strictly TD state, so 
caution must be used when interpreting these (see Section \ref{spl}). 
Of the remaining sources, GX~339$-$4 is a variable Galactic low-mass X-ray binary (LMXB) and LMC X-3 and X-1 are high mass X-ray binaries (HMXBs)
in the  Large Magellanic Cloud. Despite LMC X-3 being an HMXB, it likely accretes via Roche-lobe overflow \citep{soria_etal_2001}. 
On the other hand, LMC X-1 is thought to accrete via a stellar wind \citep{orosz_etal_2009}, but is unusual as it has a stable accretion disk. 
We note that \cite{ruhlen_and_smith_2010} have 
suggested that it may be a hybrid Roche-lobe overflow/wind accretion system.

\begin{table*}
\centering
\caption{The ULX sample}
\label{ULX_sample}
\begin{tabular}{ccccccc}
\hline
Src ID$^a$ & XMM/CXO ID$^b$ & Distance$^c$ & ${N_{\rm H}}^d$ & Obs ID$^e$ & ${\rm{t_{exp}}}^f$ & Count rate$^g$ \\
           &                & (Mpc)        &$(10^{20}~{\rm cm^{-2}})$&    & (ks)               & $({\rm s^{-1}})$ \\
\hline
M31 ULX2 & XMMU J004243.6+412519 & 0.79 & 6.85 & 0674210601 & 15.1/18.6/18.8 & 0.65/0.59/0.54 \\
M31 ULX1 & CXOM31 J004253.1$+$411422 & 0.79 & 6.68 & 0600660201 & 14.5/-/- & 6.2/-/- \\
NGC 253 XMM2 & 2XMM J004722.6$-$252050 & 3.68 & 1.38 & 0152020101 & 55.9/76.1/76.8 & 0.25/0.083/0.084 \\
NGC 253 ULX2 & 2XMM J004732.9$-$251749 & 3.68 & 1.38 & 0152020101 & 56.0/76.1/76.8 & 0.22/0.075/0.074 \\
M33 X-8 & 2XMM J013350.8$+$303937 & 0.92 & 5.69 & 0650510201 & 61.1/-/- & 4.5/-/-\\
NGC 2403 X-1 & 2XMM J073625.5$+$653540 & 3.50 & 4.17 & 0164560901 & 50.8/69.3/72.4 & 0.29/0.099/0.10 \\
NGC 4736 ULX1 & 2XMM J125048.6$+$410743 & 4.66 & 1.44 & 0404980101 & 32.7/42.9/43.3 & 0.26/0.076/0.078 \\
\hline \\
\end{tabular}
\begin{minipage}{\linewidth}
Notes:
$^a$common source name, which in the case of M31 ULX2 this has not been used previously, but it is 
defined here for convenience;
$^b$catalog identifier of the source;
$^c$distance to the source; 
$^d$Galactic neutral hydrogen column density in the direction of the ULX 
from \cite{dickey_and_lockman_1990};
$^e${\xmmn} observation identifier of the data sets used in this work, all of which were taken in full frame imaging mode;
$^f$good exposure time in ks for the EPIC pn/MOS1/MOS2 detectors, dashes indicate that data from the corresponding detector was not used in this study; 
$^g$count rate for the EPIC pn/MOS1/MOS2 detector.
\end{minipage}
\end{table*}

The main purpose of this work was to compare the most luminous TD sub-Eddington BHBs 
with a sample of BD ULXs. As such, we defined a ULX sample as those sources identified 
as having BD spectra and 0.3--10 keV X-ray luminosities $< 3 \times 10^{39}~{\rm \ergs}$ by 
\cite{sutton_etal_2013b}. To these we add another ULX in M31 (M31 ULX2), which is a transient that at its peak had an unabsorbed 
X-ray luminosity of $\sim 1.3 \times 10^{39} \ergs$ \citep{middleton_etal_2013}. 
For each source we consider only the 
observation with the highest quality data in either \cite{sutton_etal_2013b} or \cite{middleton_etal_2013}, 
details of which are given in Table \ref{ULX_sample}. 
The observations chosen for M31 ULX1 and ULX2 were the most luminous reported by these authors, and that chosen for NGC 253 XMM2 
was close to the peak luminosity. The NGC 253 ULX2 and M33 X-8 observations were below peak luminosity, and only single observations 
of both NGC 2403 X-1 and NGC 4736 ULX1 were reported in \cite{sutton_etal_2013b}. 
Out of the ULX sample three sources have been reported to be transients: M31 ULX2 \citep{barnard_etal_2013,middleton_etal_2013}, 
M31 ULX1 \citep{kaur_etal_2012,middleton_etal_2012}, and NGC 4736 ULX1 \citep{akyuz_etal_2013}. 
The other 4 ULXs are consistently detected in previous observations, so 
they are potentially persistent or long period transient sources (e.g., 
\citealt{schlegel_and_pannuti_2003,sutton_etal_2013b,sutton_etal_2014,laparola_etal_2015}). 
Given their X-ray luminosities, the ULXs are most likely powered by Roche lobe overflow. 
Faint ULXs may be either HMXBs or LMXBs \citep{swartz_etal_2004}, 
although it has been argued that ULXs 
are predominantly HMXBs 
(e.g., \citealt{king_etal_2001}).

\section{analysis}\label{analysis}

\begin{figure*}
\begin{center}
\includegraphics[width=16cm]{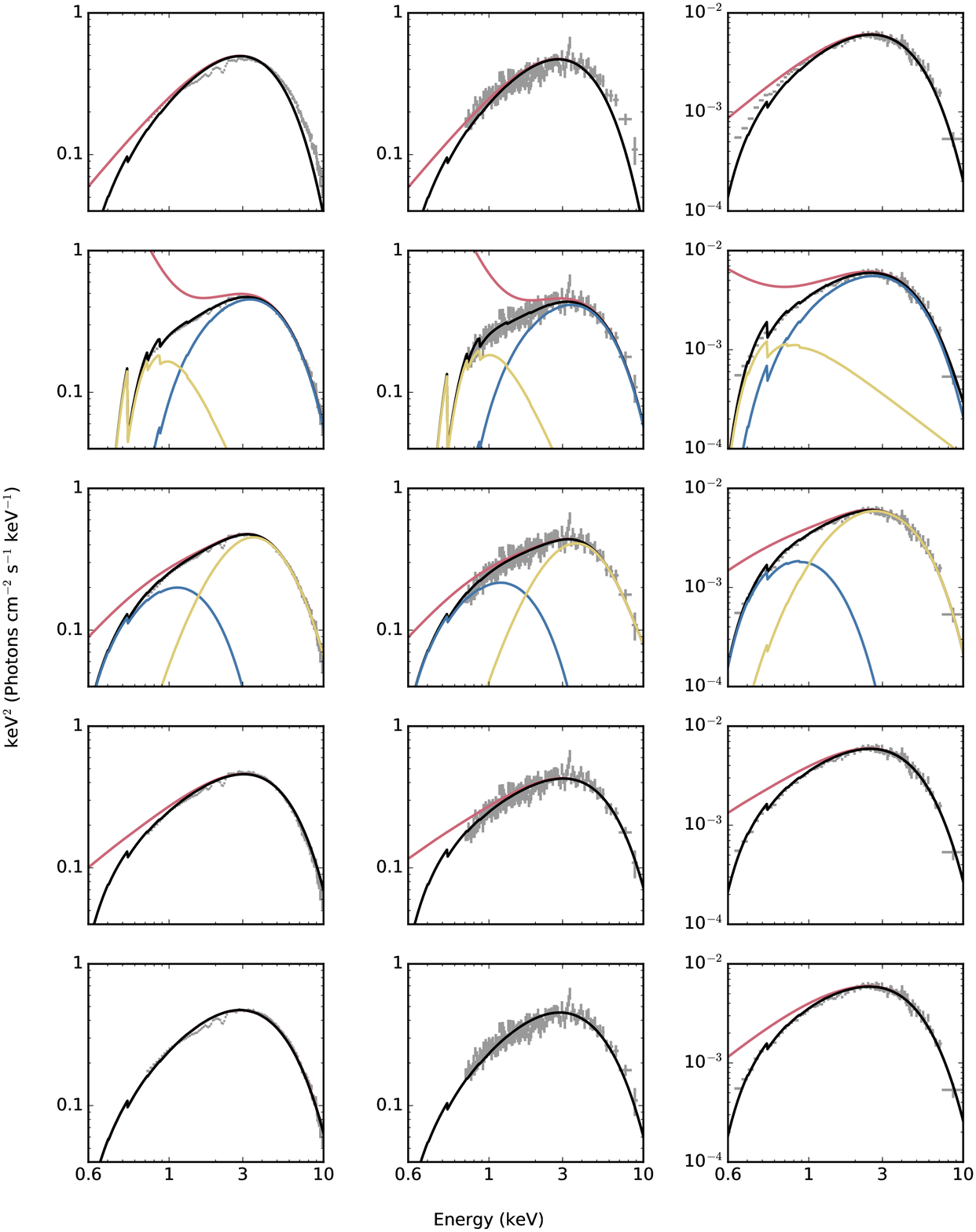}
\caption{Examples of different spectral models fitted to TD BHB and BD ULX data. The gray data points show the real (left) and resampled 
(center) spectra from LMC X-3 {\xmm}-1, and the data from (right) M31 ULX1.  
All of the spectra have been rebinned to $10 \sigma$ significance for clarity. 
The lines show the various models used in this work, which are (from top to bottom): an MCD, 
an MCD plus a power-law, an MCD plus Comptonization, a $p$-free disk, and the {\sc kerrbb} 
approximation of a thin accretion disk spectrum. 
Specifically, the lines show:
the observed model (black); the unabsorbed model (red); the MCD in two-component models (blue); and 
the power-law/Comptonization in two-component models (yellow). Clearly there is a large degree of degeneracy 
between the different spectral models.}
\label{spec_eg}
\end{center}
\end{figure*}

We extracted X-ray spectra for all of the {\xmmn} and {\swift} observations in Tables \ref{BHB_obs} 
and \ref{ULX_sample}. {\xmmn} spectra were extracted using Science Analysis Software ({\sc sas}) 
version 14.0.0.\footnote{\url{ http://xmm.esac.esa.int/sas/}} 
We considered only the EPIC pn data for the high flux TD BHBs, although EPIC pn, MOS1, and MOS2 data were all used for the lower flux ULXs.
The pn and MOS observation data files were processed using {\sc epproc} and {\sc emproc} respectively. 
Full field high energy ($>10~{\rm keV}$) 
light curves were extracted and used to define Good Time Intervals (GTIs), using {\sc tabgtigen}, to filter out periods of high 
background flaring. 

The definition of source and background regions differed between the ULXs and BHBs. 
For the BHB observations, which were in timing and burst mode, rectangular 
source regions were defined in raw CCD coordinates. These spanned 18 pixels in RAW-X centered on the source 
`streak' in the raw coordinate image. For observations in timing mode the full range of RAW-Y was used. 
For the burst mode observations pixels with RAW-Y greater than 140 were excluded 
\citep{kirsch_etal_2006}.  
Sources with count rates $\gtrsim 200~{\rm s^{-1}}$ dominate the events over the entire {\xmmn} EPIC pn 
CCD in some bands, even in timing modes.\footnote{\url{http://xmm2.esac.esa.int/docs/documents/CAL-TN-0083.pdf}} 
As a result of this, there is no uncontaminated region from which to extract a background spectrum. 
Simply approximating a background spectrum defined at the edge of the CCD would modify the source spectrum in at least some bands as the 
PSF is energy dependent \citep{done_and_diaz-trigo_2010}. Therefore, it is recommended not to subtract any background at all \citep{ng_etal_2010}. 
As such, the only {\xmmn} BHB observation for which we extracted a background spectrum is {\xmm}-1 of LMC X-1 where 
the count rate is lower than $200~{\rm s^{-1}}$. In this case, the background was defined in a region spanning RAW-X = 3--5. 
The TD BHB source and background spectra were extracted using {\sc evselect} 
with standard pattern and flag selections (${\rm PATTERN} \le 4$ and ${\rm FLAG} = 0$).
The task {\sc backscale} was used to add the BACKSCAL keyword to the headers of the source and background spectra from 
LMC X-1 to account for 
different source and background region sizes. 

For the ULX observations, which are in imaging mode, source spectra were extracted from circular regions centered on the sources, 
with typical radii of $\sim 30~{\rm arcsec}$, although smaller regions were used for NGC 253 ULX2 (10 arcsec), NGC 4736 ULX1 (20 arcsec) 
and the pn detection of NGC 253 XMM2 (25 arcsec) due to neighboring sources or proximity to chip gaps. Background spectra were extracted from larger 
circular regions (typically $\gtrsim 30~{\rm arcsec}$ for pn and $\gtrsim 50~{\rm arcsec}$ for MOS1/2). For the MOS exposures, 
these were defined in source free regions on the same chip as the source. In ideal cases, pn 
background spectra were extracted from source free regions on the same chip as the source at similar distances from the readout node. 
Where this was not possible, backgrounds were extracted from neighboring CCDs in the same quadrant as the source and at a similar distance 
from the readout node. In the pn detection of M33 X-8, following the above prescription and defining a 
background region in a source free area of a neighboring CCD in the same quadrant resulted in under-subtraction of the Cu/Ni detector line 
complex at $\sim 8$~keV. As such, for this source we compromised by extracting the background spectrum from the same CCD as the source, but at 
a slightly larger distance from the readout node. A comparison of this and a standard background defined on the neighboring chip demonstrated 
that the final spectra did not differ significantly outside of the energy range of the Cu/Ni detector lines. 
Spectra were extracted using {\sc evselect} with the recommended FLAG and PATTERN options 
(${\rm PATTERN} \le 4$ and $FLAG = 0$ for pn, and $PATTERN \le 12$ and the \#XMMEA\_EM flag selection for MOS), {\sc backscale} was used to add 
the BACKSCAL keyword to the headers of the source and background spectra. 

Redistribution Matrix Files (RMFs) and Ancillary Response Files (ARFs) were generated using 
{\sc rmfgen} and {\sc arfgen} for all of the BHB and ULX {\xmmn} observations. Source and, in cases where they 
could be extracted, background spectra were 
associated with the appropriate RMFs and ARFS then binned to a 
minimum of 20 counts using {\sc specgroup}. For the BHB observations we also set the oversample parameter to 3 to avoid oversampling the energy 
resolution of EPIC pn due to the high fluxes. During the processing of the ULX observations, a check of the event pattern 
distributions indicated that M31 ULX1 and M33 X-8 were slightly piled-up. As such, we generated 
RMFs with pile-up corrections\footnote{\url{http://xmm.esac.esa.int/sas/current/documentation/threads/\\epatplot.shtml}} for the pn data from 
these sources, and excluded the MOS exposures for which such corrections are not yet available in {\sc sas}.

{\swift} XRT spectra were extracted using {\sc xselect} from events files with default selection criteria (i.e., grade 0--2  
events for windowed timing mode). Source and background spectra were extracted from circular regions 
and the values of the BACKSCAL keyword were manually changed such that they reflected the 1-dimensional size of 
the regions. 
The appropriate RMF for windowed timing mode and grade 0--2 events was used from the calibration database, and ARFs were 
created for each observation using {\sc xrtmkarf} with exposure maps from the pipeline data reduction. Finally, spectra and 
response files were grouped and binned to a minimum of 20 counts per bin using {\sc grppha}. We do not include an additional 
3\% fractional error, as is recommended to account for systematic uncertainties in {\swift} XRT spectra, since these are not 
random errors and are likely 
correlated. As such, the systematic uncertainties in the instrument response were reflected in the value of $\chi^2$ when 
fitting these spectra.

To minimize the effects of very different signal-to-noise ratios in the nearby BHBs and relatively distant ULXs, we 
under-sampled each of the real BHB spectra to obtain realizations with $\sim 2.5 \times 10^4$ counts.  
We did this using a {\sc python} script, which 
first normalized each of the unbinned source spectra such that they contained a total of $2.5 \times 10^4$ 
counts (and, where appropriate, applied the same normalization factor to the corresponding background spectrum), then used the number of 
counts in each energy channel of the normalized spectra as expectation values to generate random numbers of counts from Poisson
distributions. The random numbers of counts were used to construct the resampled spectra. These were then binned to a minimum of 
20 counts and grouped with the corresponding resampled background spectra (where appropriate), RMFs, and ARFs using {\sc grppha}. As the 
exposure times were not altered in the header files, an additional multiplicative constant was included in the subsequent spectral fitting to 
account for the renormalization in flux. This was set equal to the ratio of the number of source counts in the simulated and real data.

Once extracted, the BHB (both real and resampled) and ULX spectra were read in to {\sc xspec} v12.8.2 \citep{arnaud_1996} and fitted with various models 
in the 0.3--10 ({\xmmn} imaging mode and {\swift} observations) or 0.7--10 keV ({\xmmn} timing and burst mode observations) 
energy ranges, 
using abundances from \cite{wilms_etal_2000} and photoionization absorption cross-sections from \cite{BCMC}, 
with a new He cross-section based on \cite{BCMC2}. 
Five spectral models were used, which 
we discuss in detail in the following section. 
For the extra-Galactic sources (i.e., all sources except 
for GX~339$-$4) an additional {\sc tbabs} absorption component was included, which was fixed to the Galactic column density in the 
direction of the source (Tables \ref{BHB_catalogue} and \ref{ULX_sample}). For fits to NGC 253 ULX2 an additional 0.67 keV {\sc mekal} component was 
included, which was subject to Galactic absorption only, to account for diffuse emission around the source (cf. \citealt{sutton_etal_2013b}). 
Further details of the specific models and results from the spectral fitting are given in Section \ref{results}.

In addition to the spectral analysis, we also carried out a timing analysis. 
We extracted covariance spectra to test 
whether they were consistent with being constant, thus could originate from single-component energy spectra. 
Covariance is advantageous here over the raw rms variability, as by design the uncorrelated white noise cancels out, which reduces the 
statistical error. 
We calculated covariances 
in the time domain \citep{wilkinson_and_uttley_2009} over time-scales of 200 s -- 7 ks. This choice of time-scale was driven by the typical ULX 
count rates and observation lengths. 
It was not possible to study variability on these time-scales in the {\swift} observations of LMC X-3 and LMC X-1 due to the orbit of the observatory, 
so these were excluded from this analysis. 
We note that the method of extracting covariance in the time domain (instead of frequency) could be affected by time lags. 
However, the time-scales that we probed were much greater than lags expected in BHBs and reported in NGC 5408 X-1 
(\citealt{heil_and_vaughan_2010, demarco_etal_2013} 
albeit in a more luminous ULX with a distinct two-component X-ray spectrum), although we caution that \cite{hernandezgarcia_etal_2015} tentatively 
identified $\sim 1~{\rm ks}$ lags below frequencies of 1 mHz at low significance in the same source.

Before covariances could be calculated, we needed to extract suitable light curves from each of the observations. 
For several of the ULXs the GTIs that we used for the spectral extraction contained many short periods of isolated good time, so we 
created new GTIs for these with binning and time limits 
chosen such that they were suitable for extracting simultaneous light curves in all of the EPIC detectors with multiple segments of 7~ks length. 
Source and, for the ULXs and LMC X-1, background light curves were extracted using {\sc evselect}. 
As we would be combining light curves from multiple detectors for the ULXs, 
fixed start and stop times were explicitly set for each observation. The light curves were corrected for 
e.g., background, dead time, and quantum efficiency using {\sc epiclccorr}. Short gaps that occurred in the corrected light curves, due to zero good 
exposure time within a 
temporal bin, were filled by averaging over neighboring bins. For the ULXs the light curves from each detector in a given 
observation were summed to produce combined EPIC light curves. 
Finally, the light curves were re-binned to 200s.
Initially, we produced light curves in 5 energy bins equally sized in log space between 0.3 and 10 keV for each observation. 
If any of the light curves for a given observation contained fewer than $\sim 20$ counts per 200 s bin on average, then the number of energy bins was 
reduced by one and the process repeated. 

For each of the combined EPIC light curves, fractional rms variability was calculated in the time domain in 7~ks continuous segments following 
\cite{vaughan_etal_2003}, 
then averaged over the multiple segments. Where variability was detected at $\ge 3 \sigma$ significance in at least one energy band, we went 
on to calculate covariance. 
The energy band with the most significant detection of fractional rms was used as the reference band. Covariances were normalized by 
$1 / \sqrt{\sigma_{\rm xs,y}^2}$ 
\citep{wilkinson_and_uttley_2009}, where $\sigma_{\rm xs,y}^2$ is the excess variance in the reference band, then divided by the mean 
count rate in the band of interest 
to obtain fractional covariances. These were again averaged over the multiple 7~ks segments. 
Where there were fewer than four light curve segments we calculated the error on the covariance following \cite{wilkinson_and_uttley_2009}, otherwise 
we used standard errors. 
Once extracted, the fractional covariance spectra were fitted with 
constants.

\section{results}\label{results}

\begin{table*}
\centering
\caption{MCD}
\label{diskbb}
\begin{tabular}{cccccc}
\hline
Source        & Obs ID$^a$ & ${N_{\rm H}}^{b}$          & ${T_{\rm in}}^{c}$ & ${R_{\rm in}}^d$ & $\chi^2/{\rm dof}^e$ \\
& &($10^{22}~{\rm cm^{-2}}$)&(keV)&(km)& \\
\hline
\multicolumn{6}{c}{Observed BHBs}\\
GX~339$-$4    & {\xmm}-1   & 0.62                       & 0.91 & 43 & (8222.6/163)  \\
              & {\xmm}-2   & 0.61                       & 0.92 & 42 & (3556.9/160)  \\  
LMC X-3       & {\xmm}-1   & 0                          & 1.2  & 57 & (28994.7/164) \\  
              & {\swift}-1 & 0                          & 1.1  & 54 & (2787.3/616)  \\  
              & {\xmm}-2   & 0                          & 1.1  & 56 & (18379.2/162) \\  
LMC X-1       & {\xmm}-1   & 0.56                       & 0.86 & 51 & (1932.9/151)  \\  
              & {\swift}-1 & $0.491 \pm 0.005$ & $0.952 \pm 0.004$ & 44 & (818.2/564) \\  
              & {\swift}-2 & $0.455 \pm 0.007$ & $0.988 \pm 0.006$ & 39 & (684.3/519) \\

\hline
\multicolumn{6}{c}{Resampled BHBs}\\
GX~339$-$4    & {\xmm}-1   & $0.63 \pm 0.01$           & $0.896 \pm 0.008$ & 44 & 593.9/559   \\
              & {\xmm}-2   & $0.61 \pm 0.01$           & $0.905 \pm 0.009$ & 43 & 606.1/566   \\  
LMC X-3       & {\xmm}-1   & [0]                       & $1.18 \pm 0.01$   & 59 & (689.2/502) \\ 
              & {\swift}-1 & [0]                       & $1.08 \pm 0.01$   & 54 & (657.1/393) \\  
              & {\xmm}-2   & [0]                       & $1.05 \pm 0.01$   & 56 & (711.1/481) \\  
LMC X-1       & {\xmm}-1   & $0.57 \pm 0.01$           & $0.849 \pm 0.008$ & 51 & (687.7/556) \\  
              & {\swift}-1 & $0.366 \pm 0.009$         & $0.918 \pm 0.009$ & 48 & (514.0/391) \\  
              & {\swift}-2 & $0.331^{+0.009}_{-0.008}$ & $0.98 \pm 0.01$   & 40 & (576.9/396) \\
\hline
\multicolumn{6}{c}{ULXs}\\
M31 ULX2      & \ldots     & $0.296 \pm 0.007$              & $0.858 \pm 0.008$ & $> 110$ & (528.5/309)  \\  
M31 ULX1      & \ldots     & 0                              & 0.99              & $>  93$ & (678.7/138)  \\  
NGC 253 XMM2  & \ldots     & $0.044 \pm 0.004$              & $1.16 \pm 0.01$   & $>  72$ & (453.5/326)  \\ 
NGC 253 ULX2  & \ldots     & $0.22 \pm 0.01$                & $1.66 \pm 0.02$   & $>  46$ & 331.9/347    \\  
M33 X-8       & \ldots     & 0                              & 1.1               & $>  84$ & (1849.0/163) \\  
NGC 2403 X-1  & \ldots     & $0.162^{+0.006}_{-0.005}$      & $1.04 \pm 0.01$   & $> 100$ & 371.3/330    \\  
NGC 4736 ULX1 & \ldots     & 0                              & 0.80              & $> 160$ & (608.0/241)  \\  
\hline \\
\end{tabular}
\begin{minipage}{\linewidth}
Parameters and fit statistics for the absorbed MCD model 
($\textsc{tbabs} \times \textsc{diskbb}$ in {\sc xspec}) fitted to the the observed 
and resampled BHB spectra and the ULX spectra. Errors and limits are given to $1 \sigma$ significance, and unconstrained parameters 
are highlighted in square brackets. 
Parameters are arbitrarily quoted to two significant figures 
for observations where the fit statistic $(\chi^2/{\rm dof})$ exceeds a value of two. 
For each of the extra-Galactic sources an additional absorption component ({\sc tbabs}) was included to account for 
the Galactic column density, which was fixed to the values in Tables \ref{BHB_catalogue} and \ref{ULX_sample}. 
Notes:
$^a$observation identifier for the BHBs (see Table \ref{BHB_obs});
$^b$absorption column, 
(values below $1 \times 10^{-3}$ are shown as 0);
$^c$inner disk temperature of the MCD; 
$^d$approximate inner disk radius calculated using the best-fitting model parameters and arbitrarily reported to two significant figures, 
it is corrected from the apparent radius using $R_{\rm in} \sim 1.19 \times r_{\rm in}$ {\citep{kubota_etal_1998}}, 
inclinations and distances from Table \ref{BHB_catalogue} were assumed for the BHBs 
($i \sim 25 \deg$ was assumed for GX~339$-$4), and distances from Table \ref{ULX_sample} were used for the ULXs, where the lack of 
inclination estimates meant only a lower-limit can be given; 
$^e$goodness of fit statistic in terms of $\chi^2/{\rm degrees~of~freedom}$, where values in 
brackets indicate that the model can be rejected at the equivalent of $\ge 3 \sigma$ significance.
\end{minipage}
\end{table*}

\begin{table*}
\centering
\caption{MCD plus power-law}
\label{diskbb_pl}
\begin{tabular}{cccccccc}
\hline
Source & Obs ID & ${N_{\rm H}}$ & ${T_{\rm in}}^{a}$ & $\Gamma^b$ & ${F_{\rm pow}/F_{\rm diskbb}}^c$ & ${R_{\rm in}}$ & $\chi^2/{\rm dof}$ \\
 & & $(10^{22}~{\rm cm^{-2}})$ & (keV) & & & (km) & \\
\hline
\multicolumn{8}{c}{Observed BHBs}\\
GX~339$-$4    & {\xmm}-1   & 0.92                       & 0.90                      & 3.2                  & 1.4                    & 40 & (520.9/161) \\
              & {\xmm}-2   & $0.880^{+0.010}_{-0.009}$  & $0.896 \pm 0.002$         & $3.03 \pm 0.03$      & $1.23^{+0.04}_{-0.05}$ & 40 & (289.2/158) \\
LMC X-3       & {\xmm}-1   & 0.48                       & 1.3                       & 4.6                  & 5.0                    & 46 & (2325.3/162) \\
              & {\swift}-1 & $0.011^{+0.006}_{-0.005}$  & $1.178 \pm 0.008$         & $2.27 \pm 0.07$      & $0.61 \pm 0.04$        & 42 & (899.2/614) \\
              & {\xmm}-2   & 0.33                       & 1.2                       & 3.9                  & 3.3                    & 40 & (492.1/160) \\
LMC X-1       & {\xmm}-1   & 1.2                        & 0.90                      & 4.1                  & 4.5                    & 41 & (452.5/149) \\
              & {\swift}-1 & $0.501^{+0.006}_{-0.005}$  & $0.934^{+0.006}_{-0.010}$ & $-1 \pm 1$           & $<7 \times 10^{-4}$    & 47 & (759.0/562) \\
              & {\swift}-2 & $0.48^{+0.02}_{-0.01}$     & $0.92^{+0.02}_{-0.01}$    & $1.1^{+0.6}_{-0.9}$  & $< 0.1$                & 44 & 602.9/517 \\
\hline

\multicolumn{8}{c}{Resampled BHBs}\\
GX~339$-$4    & {\xmm}-1   & $1.0 \pm 0.1$              & $0.90 \pm 0.02$           & $3.6 \pm 0.4$        & $1.8^{+0.7}_{-0.5}$    & 41 & 579.3/557 \\
              & {\xmm}-2   & $0.80^{+0.08}_{-0.09}$     & $0.85 \pm 0.02$           & $2.6^{+0.3}_{-0.4}$  & $0.8^{+0.5}_{-0.3}$    & 44 & 561.6/564 \\
LMC X-3       & {\xmm}-1   & $0.5 \pm 0.1$              & $1.37^{+0.04}_{-0.03}$    & $4.6^{+0.4}_{-0.5}$  & $6 \pm 2$              & 42 & 480.1/500 \\
              & {\swift}-1 & $<9 \times 10^{-3}$        & $1.164^{+0.017}_{-0.009}$ & $2.2^{+0.2}_{-0.1}$  & $0.50^{+0.06}_{-0.04}$ & 43 & 435.5/391 \\
              & {\xmm}-2   & [0]                        & $1.16^{+0.02}_{-0.03}$    & $2.5^{+0.2}_{-0.1}$  & $3 \pm 1$              & 41 & 527.5/479 \\
LMC X-1       & {\xmm}-1   & $1.3 \pm 0.2$              & $0.90^{+0.01}_{-0.02}$    & $4.4^{+0.7}_{-0.6}$  & $5 \pm 2$              & 41 & 623.9/554 \\
              & {\swift}-1 & $0.8^{+0.1}_{-0.2}$        & $0.94^{+0.02}_{-0.03}$    & $4.7^{+0.8}_{-1.1}$  & $5^{+5}_{-2}$          & 44 & (503.4/389) \\
              & {\swift}-2 & $0.36^{+0.03}_{-0.02}$     & $0.91 \pm 0.02$           & $1.5^{+0.5}_{-1.1}$  & $<0.3$                 & 45 & (546.1/394) \\
\hline

\multicolumn{8}{c}{ULXs}\\
M31 ULX2      & \ldots     & $0.50 \pm 0.04$            & $0.73^{+0.03}_{-0.02}$ & $2.4 \pm 0.1$        & $1.6 \pm 0.4$          & $> 110$ & 308.5/307 \\
M31 ULX1      & \ldots     & $0.03 \pm 0.02$ & $1.080 \pm 0.007$      & $3.2 \pm 0.2$        & $1.01 \pm 0.09$        & $>  74$ & 179.6/136 \\
NGC 253 XMM2  & \ldots     & $0.17 \pm 0.02$            & $1.15^{+0.07}_{-0.06}$ & $2.1 \pm 0.1$        & $1.9 \pm 0.3$          & $>  56$ & 321.9/324 \\
NGC 253 ULX2  & \ldots     & $0.26^{+0.12}_{-0.04}$     & $1.64^{+0.08}_{-0.09}$ & $2 \pm 1$            & $0.3^{+0.5}_{-0.2}$    & $>  74$ & (314.5/161) \\ 
NGC 2403 X-1  & \ldots     & $0.45^{+0.06}_{-0.05}$     & $1.13 \pm 0.02$        & $3.3^{+0.4}_{-0.3}$  & $2.3 \pm 0.3$          & $>  79$ & 310.2/328 \\
NGC 4736 ULX1 & \ldots     & $0.10 \pm 0.02$            & $1.02 \pm 0.05$        & $2.6 \pm 0.1$        & $3.1 \pm 0.4$          & $>  74$ & (315.3/239) \\
\hline \\
\end{tabular}
\begin{minipage}{\linewidth}
As Table \ref{diskbb}, but for the absorbed MCD plus power-law model 
($\textsc{tbabs} \times (\textsc{diskbb} + \textsc{powerlaw})$ in {\sc xspec}).
Notes:
$^a$inner disk temperature of the MCD;
$^b$power-law spectral index;
$^c$ratio of the 0.3--1 keV power-law and MCD fluxes calculated using the {\sc cflux} convolution model 
in {\sc xspec}, this is not a parameter of the model, but it is used to constrain the spectral type in the 
{\cite{sutton_etal_2013b}} ULX classification scheme.
\end{minipage}
\end{table*}

Various spectral models were fitted to the TD BHB and BD ULX data. These included models of pure thin accretion disks as well as 
physically motivated and phenomenological models of BDs. Examples of these 
are shown in Figures \ref{spec_eg}, and details follow. 
Uncertainties are quoted at the equivalent of $1\sigma$ significance throughout, and we reject models where 
the null hypothesis probability is greater than 0.997 (equivalent to $3 \sigma$ significance).

\subsection{MCD}

The first model used to fit the TD BHB and BD ULX X-ray spectra was of an absorbed MCD, which is an approximation 
of a thin accretion disk with only 2 free parameters: disk temperature and normalization. Fit statistics and model parameters 
are reported in Table \ref{diskbb}. A thin disk alone was rejected in the majority of BHB and ULX observations, with 
the exceptions of sub-sampled {\xmm}-1 and -2 of GX~339$-$4, NGC 253 ULX2, and NGC 2403 X-1. Although the MCD model 
could not be statistically rejected for the sub-sampled GX~339$-$4 observations, alternative two-component 
models and a broad 
disk model ($\textsc{diskpbb}$ in {\sc xspec}) 
resulted in improved fit statistics (see below).

\subsection{MCD plus power-law}

The second model was of an absorbed MCD plus a power-law. This model has two additional parameters compared to the MCD alone: 
the power-law spectral index and its normalization. \cite{sutton_etal_2013b} used this model to classify ULXs in to 
the three spectral types. They defined BD ULXs 
as having inner disk temperatures of greater than 0.5 keV and 0.3--1 keV power-law -- MCD flux ratios 
of less than 5. We emphasize that this model is intended to be phenomenological in the BD ULXs, with the power-law 
broadening the disk spectrum rather than fitting to a physically distinct component. The model parameters, fit 
statistics, and 0.3--1 keV component flux ratios for the observed BHB, resampled BHB, and ULX spectra are given in 
Table \ref{diskbb_pl}. The fit statistics indicate that the model was insufficient to reproduce the complex 
spectra from the complete BHB data, except in LMC X-1 {\swift}-2, which had some of the lowest quality data amongst the 
complete BHB observations. The model could not be rejected for 6/8 of the resampled BHBs and 5/7 ULXs. Notably, the model 
was rejected for the resampled LMC X-1 {\swift}-2 spectrum, but it was marginally acceptable for the real data from this 
source. This is likely due to statistical effects when resampling the data.

All of the sources had 
best fitting inner disk temperatures greater than 0.5 keV. Also, in all sources where errors were estimated cooler accretion disks 
were rejected at $> 3 \sigma$ significance. 
All of the spectra also had 0.3--1 keV observed power-law -- MCD flux ratios that were at least consistent with being less than 
or equal to 5, with greater values being strongly rejected ($>3 \sigma$ significance) for 4 of the resampled BHB spectra and 
all of the ULXs. As such, we were rather predictably able to confirm that all of the ULXs selected 
for having BD spectra indeed have spectra that are broader than an MCD. But, crucially, 
this spectral model also indicated that some of the brightest TD BHBs had BD-shaped spectra at X-ray luminosities an 
order of magnitude lower than the ULXs.

Interestingly, although they are only approximate, the inner disk radii calculated from 
the model normalizations appear to 
be a factor $\sim 2$ greater in the ULXs than the BHBs. If this corresponds to the inner-most stable circular orbit, 
then it would indicate that the black holes in the BD ULXs were larger than in these TD BHBs.

\begin{table*}
\centering
\caption{MCD plus Comptonization}
\label{diskbb_comptt}
\begin{tabular}{ccccccccc}
\hline
Source & Obs ID & ${N_{\rm H}}$ & ${T_{\rm in}}^{a}$ & ${T_{0}}^b$ & $kT^c$ & $\tau^d$ & ${R_{\rm in}}$ & $\chi^2/{\rm dof}$ \\
 & & $(10^{22}~{\rm cm^{-2}})$ & (keV) & (keV) & (keV) & & (km) & \\
\hline
 \multicolumn{9}{c}{Observed BHBs}\\
GX~339$-$4 & {\xmm}-1   & 0.76 & 0.39 & 0.59 & 4.5  & 1.1 & 180 & (603.9/159) \\
           & {\xmm}-2   & $0.742 \pm 0.008$  & $0.40 \pm 0.01$ & $0.590^{+0.009}_{-0.010}$ & $2.3^{+15.1}_{-0.4}$ & $2.7^{+0.7}_{-2.4}$ & 170 & (300.5/156) \\
LMC X-3    & {\xmm}-1   & 0    & 0.44 & 0.72 & 1.4  & 6.6 & 280 & (1929.1/160) \\
           & {\swift}-1 & [0] & $0.350 \pm 0.007$ & $0.62^{+0.01}_{-0.02}$ & $1.9^{+1.5}_{-0.3}$ & $4^{+1}_{-2}$ & 320 & (726.7/612) \\
           & {\xmm}-2   & [0] & $0.405 \pm 0.006$ & $0.63 \pm 0.02$ & $1.30 \pm 0.04$ & $7.1^{+0.4}_{-0.5}$ & 260 & (315.1/158) \\
LMC X-1    & {\xmm}-1   & 0.96 & 0.27 & 0.51 & 1.2  & 5.4 & 580 & (351.6/147) \\
           & {\swift}-1 & $0.65 \pm 0.03$ & $0.28 \pm 0.02$ & $0.52 \pm 0.01$ & $7^{+31}_{-6}$ & $<3$ & 410 & (730.4/560) \\
           &  {\swift}-2 & $0.61 \pm 0.04$ & $0.29^{+0.03}_{-0.02}$ & $0.51^{+0.02}_{-0.01}$ & $<70$  & $<9$ & 340 & 594.8/515 \\
\hline

\multicolumn{9}{c}{Resampled BHBs}\\
GX~339$-$4 & {\xmm}-1   & $0.57^{+0.06}_{-0.05}$ & [$4 \times 10^{-3}$] & $0.36 \pm 0.04$ & $1.08 \pm 0.03$ & $10.8 \pm 0.6$    & $2.3 \times 10^4$ & 577.3/555 \\
           & {\xmm}-2   & $0.8 \pm 0.2$ & $0.23^{+0.16}_{-0.05}$ & $0.41^{+0.10}_{-0.05}$ & $1.2^{+0.3}_{-0.1}$ & $6^{+1}_{-2}$ & 580 & 557.2/562 \\
LMC X-3    & {\xmm}-1   & [0] & $0.47^{+0.04}_{-0.03}$ & $0.80 \pm 0.07$ & $4^{+63}_{-3}$ & $<9$                                & 260 & 480.1/498  \\ 
           & {\swift}-1 & [0] & $0.35 \pm 0.02$ & $0.63 \pm 0.03$ & $5^{+72}_{-3}$ & $<7$                                       & 320 & 416.7/389 \\
           & {\xmm}-2   & $<0.4$ & $0.19^{+0.06}_{-0.05}$ & $0.33^{+0.07}_{-0.11}$ & $1.12^{+0.07}_{-0.05}$ & $10 \pm 1$        & $1.8 \times 10^3$ & 522.3/477 \\
LMC X-1    & {\xmm}-1   & $1.03 \pm 0.09$ & $0.25^{+0.03}_{-0.02}$ & $0.50^{+0.03}_{-0.02}$ & $<40$ & $<4$                      & 770 & 614.5/552 \\
           & {\swift}-1 & $0.67^{+0.07}_{-0.08}$ & $0.22^{+0.04}_{-0.03}$ & $0.46 \pm 0.06$ & $<9$ & $6^{+2}_{-5}$              & $1.1 \times 10^3$ & (483.0/387) \\
           & {\swift}-2 & $0.70 \pm 0.06$ & $0.21 \pm 0.02$ & $0.47 \pm 0.02$ & $50^{+20}_{-40}$ & $<5$                         & $1.5 \times 10^3$ & (509.9/392) \\
\hline

\multicolumn{9}{c}{ULXs}\\
M31 ULX2      & \ldots & $0.37 \pm 0.03$                    & $0.65^{+0.13}_{-0.02}$     & $1.34^{+1.44}_{-0.09}$ & $<100$                 & $<60$                  & $> 180$ & 312.4/305 \\
M31 ULX1      & \ldots & $0.015^{+0.008}_{-0.010}$          & $0.30^{+0.04}_{-0.03}$     & $0.44^{+0.05}_{-0.03}$ & $<1.03$                & $10.7^{+0.2}_{-0.8}$   & $> 610$ & 176.0/134 \\
NGC 253 XMM2  & \ldots & $0.27 \pm 0.03$                    & $0.07 \pm 0.01$            & $<0.2$                 & $1.34^{+0.08}_{-0.06}$ & $9.0 \pm 0.5$          & $> 3.7 \times 10^4$ & 313.6/322 \\
NGC 253 ULX2  & \ldots & $0.3^{+0.3}_{-0.2}$                & $0.10^{+0.03}_{-0.02}$     & $0.29^{+0.05}_{-0.28}$ & $1.41 \pm 0.05$        & $11.2^{+0.6}_{-1.3}$   & $> 1.0 \times 10^4$ & 318.6/343 \\
M33 X-8       & \ldots & $0.047 \pm 0.003$                 & $0.48^{+0.03}_{-0.02}$     & $0.65 \pm 0.03$        & $<100$                 & $<3$                   & $> 310$ & (302.0/159) \\
NGC 2403 X-1  & \ldots & $0.26 \pm 0.02$                    & $0.41^{+0.05}_{-0.08}$     & $0.72^{+0.09}_{-0.25}$ & $2^{+23}_{-1}$         & $3^{+9}_{-1}$          & $> 500$ & 303.4/326 \\
NGC 4736 ULX1 & \ldots & [0]                                & $0.04 \pm 0.02$            & $0.157 \pm 0.007$      & $1.05 \pm 0.04$        & $10.5 \pm 0.4$         & $> 1.2 \times 10^5$ & 269.1/237 \\
\hline \\
\end{tabular}
\begin{minipage}{\linewidth}
As Table \ref{diskbb}, but for the absorbed MCD plus Comptonization model
($\textsc{tbabs} \times (\textsc{diskbb} + \textsc{comptt})$ in {\sc xspec}). 
Fits with hot accretion disk and cool, soft Comptonization ($T_{\rm in} >  T_0$) components, or 
low temperature coronae ($kT < T_0$) were rejected as unphysical, and 
alternative local minima in $\chi^2$ were used instead. 
Notes: 
$^a$inner disk temperature of the MCD;
$^b$input soft photon (Wien) temperature, this was not fixed to the inner disk temperature;
$^c$plasma temperature, this was constrained to be $> 1~{\rm keV}$;
$^d$plasma optical depth.
\end{minipage}
\end{table*}

\begin{figure*}
\begin{center}
\includegraphics[width=14cm]{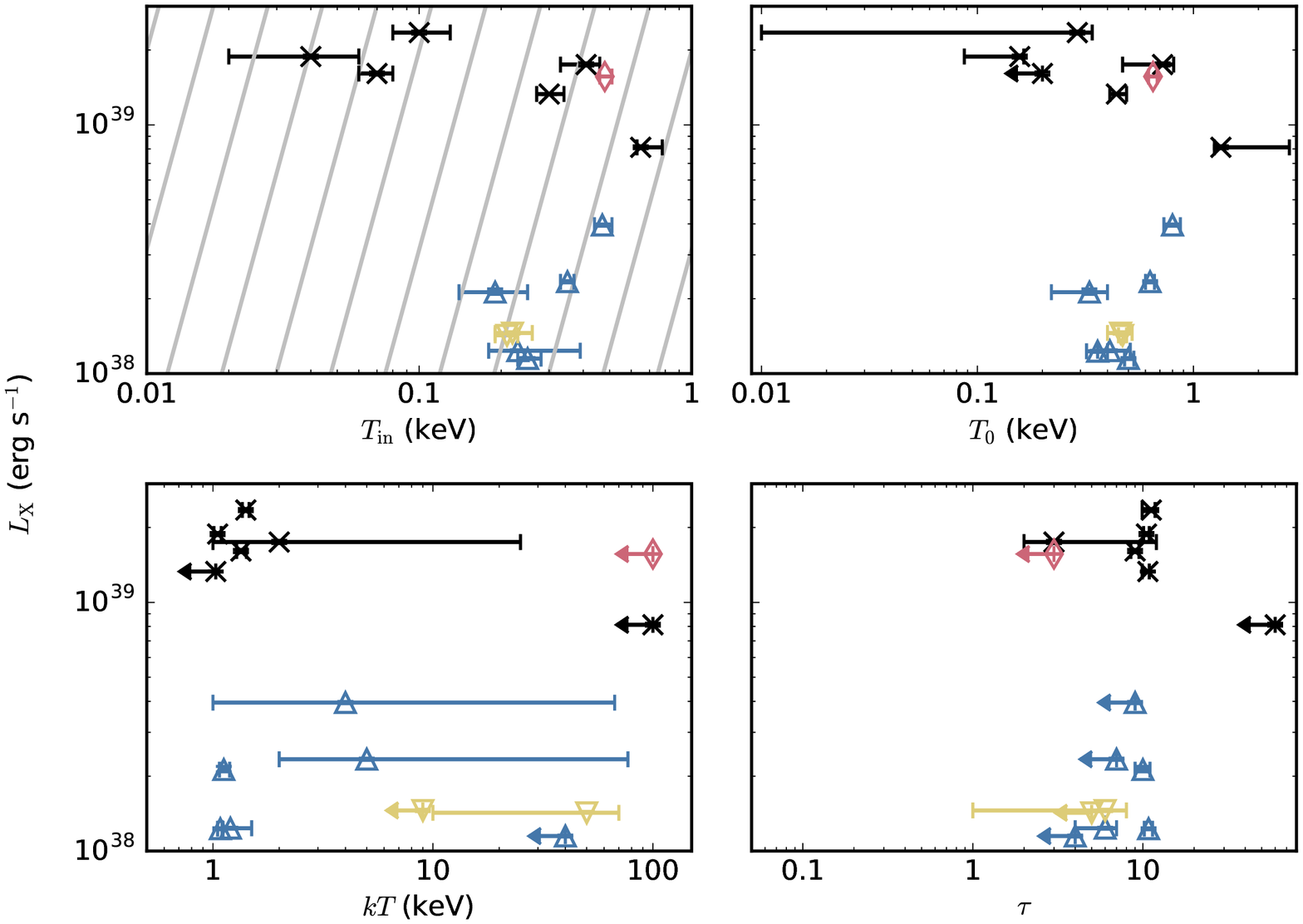}
\caption{Parameters of the absorbed MCD plus Comptonization spectral model plotted against the 
observed 0.3--10 keV luminosity estimated from the same model. 
The symbols correspond to: ULXs where the model is statistically acceptable (at $3 \sigma$ significance, black crosses), 
ULXs where the model is rejected (red diamonds), BHBs where the model is acceptable (blue upwards pointing triangles), and 
BHBs where the model is rejected (yellow downward pointing triangles).
Reading across the rows from top left, the model parameters shown are: the inner disk temperature of the MCD, the input soft 
photon temperature, the plasma temperature, and the plasma optical depth. The gray lines in top-left plot show arbitrary 
$L \propto T^4$ relations as would be appropriate for a thin accretion disk. No significant 
correlations were found between the model parameters and luminosity.}
\label{mcdcompfig}
\end{center}
\end{figure*}

\subsection{MCD plus Comptonization}

The next model fitted to the TD and BD spectra was of an MCD with additional Comptonized emission ($\textsc{diskbb} + \textsc{comptt}$ 
in {\sc xspec}). 
The free parameters of this model are: the inner accretion disk temperature, disk normalization, seed photon temperature of the Comptonization, 
plasma temperature, plasma optical depth and normalization of the Comptonized component. 
This, and similar models, have commonly been used to model two-component {\xmmn} spectra from more 
luminous ULXs (e.g., \citealt{feng_and_kaaret_2009, gladstone_etal_2009}). Generally this model results in a soft MCD and hard, 
optically thick Comptonization. When used for ULXs the Comptonization input photon temperature is often set equal to the inner 
disk temperature of the MCD. This is intended as an approximation of a Comptonizing corona above an accretion disk, but fixing 
the temperatures in this way may not be appropriate if the corona is optically thick and obscures the inner disk from view 
\citep{gladstone_etal_2009}. As such, we did not fix the Comptonization input photon temperature here. This model has also been used to 
fit BD ULX X-ray spectra, which has been physically interpreted as 
the onset of a radiatively driven wind at accretion rates close to Eddington 
(\citealt{middleton_etal_2011b, middleton_etal_2012, middleton_etal_2013}). The same model was used to fit broad, 
sub-Eddington accretion disk spectra from GX~339$-$4 when it was in the TD state \citep{kolehmainen_etal_2011}, in which case it is a 
purely phenomenological model.

The parameters and fit statistics for the MCD plus Comptonization model are given in Table \ref{diskbb_comptt}. Fits 
with input soft photon temperatures lower than the inner accretion disk temperature, or plasma temperatures lower than the input photon temperature, 
were rejected as unphysical. 
Similarly to the 
previous model, this was not sufficient to reproduce all of the complexities in most of the complete BHB spectra, although it was sufficient 
for 6/8 of the resampled BHBs and 6/7 of the ULXs. Several of the the resampled TD BHB and BD ULX spectra were best fitted by 
optically thick coronae, although many admittedly have poorly constrained model parameters. This is reminiscent of many more-luminous 
ULXs and is consistent with an optically 
thick Comptonizing corona obscuring the inner disk. 

In Figure \ref{mcdcompfig} 
we show a comparison of the $\textsc{diskbb} + \textsc{comptt}$ model parameters in the BHBs and ULXs. In general, the BHBs and ULXs 
tended to occupy similar regions of parameter space, with the exception of $L_{\rm X}$. 
After excluding the rejected fits and the unconstrained inner disk temperature in GX~339$-$4 {\xmm}-1, 
we tested for rank correlations between the parameters and $L_{\rm X}$ using 
Kendall's $\tau$ coefficient.
The resulting Kendall's $\tau$ correlation coefficients were: 0.2, 0.1, 0.2, and 0.3, with 
$p$-values (for the null hypothesis of Kendall's $\tau=0$, i.e., no correlation) were: 0.5, 0.6, 0.5, and 0.2 for parameters 
$T_{\rm in}$, $T_{0}$, $kT$, and $\tau$, none of which are significant 
even at the $2 \sigma$ level.

\begin{table*}
\centering
\caption{$p$-free accretion disk}
\label{diskpbb}
\begin{tabular}{ccccccc}
\hline
Source & Obs ID & ${N_{\rm H}}$ & ${T_{\rm in}}^{a}$ & $p^b$ & ${R_{\rm in}}^c$ & $\chi^2/{\rm dof}$ \\
 & & $(10^{22}~{\rm cm^{-2}})$ & (keV) & & (km) & \\
\hline
\multicolumn{7}{c}{Observed BHBs}\\
GX~339$-$4 & {\xmm}-1   & 0.80 & 1.0  & 0.58 &  67 & (2158.9/162) \\
           & {\xmm}-2   & 0.81 & 1.0  & 0.57 &  62 & (1080.8/159) \\
LMC X-3    & {\xmm}-1   & 0    & 1.5  & 0.64 &  83 & (3771.5/163) \\
           & {\swift}-1 & [0] & $1.321 \pm 0.008$ & $0.642 \pm 0.002$ &  78 & (920.5/615) \\
           & {\xmm}-2   & 0    & 1.3  & 0.62 &  75 & (695.1/161) \\
LMC X-1    & {\xmm}-1   & 0.85 & 1.0  & 0.51 &  58 & (581.0/150) \\
           & {\swift}-1 & $0.53 \pm 0.01$ & $0.98 \pm 0.01$ & $0.69 \pm 0.02$ & 100 & (808.1/563) \\
           & {\swift}-2 & $0.57 \pm 0.02$ & $1.08 \pm 0.02$ & $0.62 \pm 0.02$ &  69 & (649.8/518) \\
\hline

\multicolumn{7}{c}{Resampled BHBs}\\
GX~339$-$4 & {\xmm}-1   & $0.79 \pm 0.04$            & $0.97 \pm 0.02$        & $0.60^{+0.03}_{-0.02}$    & 76 & 579.0/558 \\
           & {\xmm}-2   & $0.83 \pm 0.04$            & $1.04 \pm 0.03$        & $0.55 \pm 0.02$           & 59 & 575.0/565 \\
LMC X-3    & {\xmm}-1   & $(9 \pm 7) \times 10^{-3}$ & $1.52 \pm 0.04$        & $0.613 \pm 0.009$         & 70 & 566.7/579 \\
           & {\swift}-1 & [0]                        & $1.31 \pm 0.02$        & $0.642 \pm 0.005$         & 79 & 441.4/392 \\
           & {\xmm}-2   & [0]                        & $1.32 \pm 0.03$        & $0.609^{+0.007}_{-0.006}$ & 69 & 525.5/480 \\
LMC X-1    & {\xmm}-1   & $0.86^{+0.03}_{-0.04}$     & $1.01 \pm 0.03$        & $<0.53$                    & 57 & 631.8/555 \\
           & {\swift}-1 & $0.47 \pm 0.03$            & $1.01 \pm 0.03$        & $0.59^{+0.03}_{-0.02}$    & 79 & (499.3/390) \\
           & {\swift}-2 & $0.45 \pm 0.02$            & $1.13^{+0.04}_{-0.03}$ & $0.57 \pm 0.02$           & 57 & (551.6/395) \\
\hline

\multicolumn{7}{c}{ULXs}\\
M31 ULX2      & \ldots & $0.527^{+0.006}_{-0.007}$      & $1.18 \pm 0.02$        & $<0.502$                  & $>  81$ & 351.6/308 \\
M31 ULX1      & \ldots & $<2 \times 10^{-3}$            & $1.139 \pm 0.009$      & $0.664^{+0.002}_{-0.004}$ & $> 160$ & (194.9/137) \\
NGC 253 XMM2  & \ldots & $0.15 \pm 0.01$                & $1.57^{+0.06}_{-0.05}$ & $0.576 \pm 0.009$         & $>  72$ & 326.0/325 \\
NGC 253 ULX2  & \ldots & $0.25 \pm 0.02$                & $1.75 \pm 0.05$        & $0.70 \pm 0.02$           & $> 100$ & 327.5/346 \\
M33 X-8       & \ldots & 0.099                          & 1.4                    & 0.58                      & $>  90$ & (470.7/162) \\
NGC 2403 X-1  & \ldots & $0.28 \pm 0.02$                & $1.25^{+0.04}_{-0.03}$ & $0.59 \pm 0.01$           & $> 140$ & 312.5/329 \\
NGC 4736 ULX1 & \ldots & $0.054 \pm 0.009$              & $1.31^{+0.06}_{-0.05}$ & $0.537^{+0.010}_{-0.009}$ & $>  92$ & 294.9/240 \\
\hline \\
\end{tabular}
\begin{minipage}{\linewidth}
As Table \ref{diskbb}, but for the absorbed $p$-free disk model 
($\textsc{tbabs} \times \textsc{diskpbb}$ in {\sc xspec}). 
Notes: 
$^a$inner disk temperature of the $p$-free disk in keV;
$^b$exponent of the radial dependence of the disk temperature; 
$^c$the inner disk radius is corrected from the apparent radius using 
$R_{\rm in} \sim 3.19 \times r_{\rm in}$ {\citep{vierdayanti_etal_2008}}.
\end{minipage}
\end{table*}

\begin{figure*}
\begin{center}
\includegraphics[width=14cm]{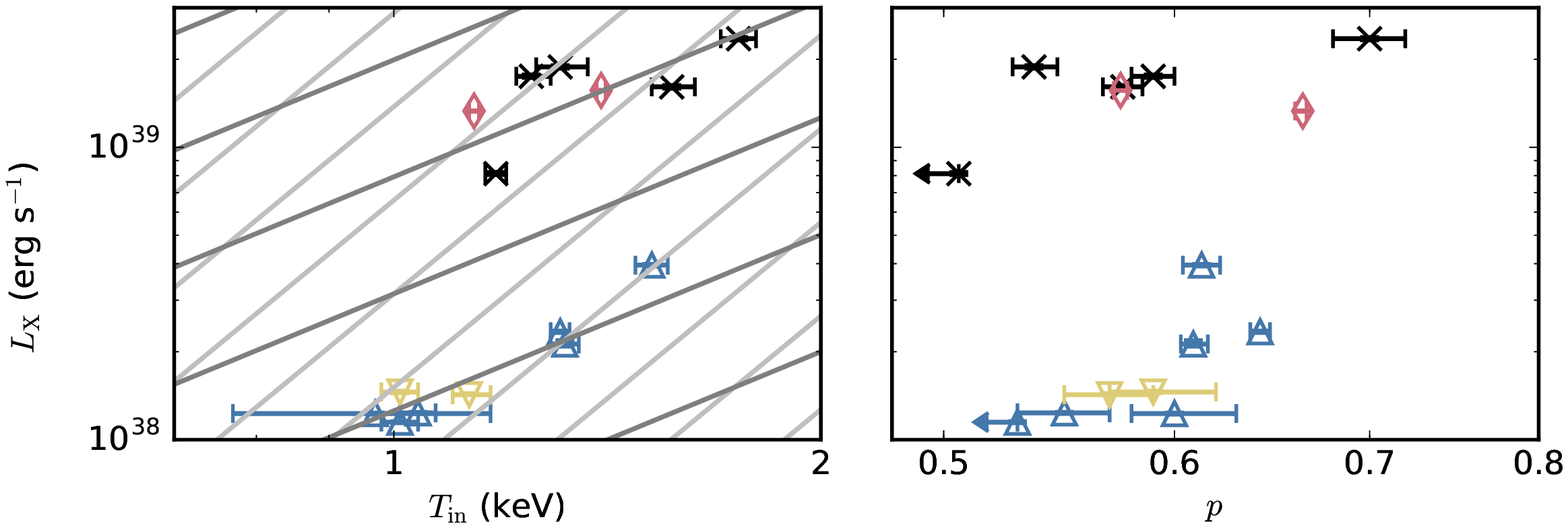}
\caption{As Figure \ref{mcdcompfig}, but for the parameters of the $p$-free disk model. The left plot shows the inner 
disk temperature, where the gray lines show characteristic luminosity -- temperature relations with arbitrary 
normalizations. The light gray lines correspond to $L \propto T^4$, which is the standard relation for thin accretion disks, and the 
dark gray lines show $L \propto T^2$, which may be appropriate for the slim disk regime \citep{kubota_and_makishima_2004}. 
The right plot shows the $p$-value. Although there is a potential correlation between luminosity and disk temperature, 
this is not highly significant (at the $3\sigma$ level).}
\label{pfreefig}
\end{center}
\end{figure*}

\subsection{p-free disk}

The fourth model was of a $p$-free disk, which is an approximation of an MCD spectrum that has been modified by radial advection. 
This model has the parameters of the MCD model (disk temperature and normalization) plus `$p$', which is the exponent of 
the radial dependence of the disk temperature 
(i.e., $T(r) \propto r^{-p}$, where $T(r)$ is the local effective temperature at radius $r$). 
A standard MCD is recovered for $p=0.75$ 
and lower values indicate a degree of advection, with $p=0.5$ being predicted for advection-dominated disks. 
Rather unsurprisingly, this model with only three free parameters (plus absorption) was unable 
to reproduce any of the real TD BHB spectra, but it was sufficient for 6/8 of the resampled BHBs and 5/7 of the ULXs. 
Notably, all of the best-fitting values of $p$ were less than 0.75 indicating that advection may have been occurring in the accretion disks. 
Furtermore, critically, $p \ge 0.75$ can be ruled out at 
$3 \sigma$ significance for all of the BHBs and 
all of the ULXs except for NGC 253 ULX2. As for the previous model we show the parameters against observed 0.3--10 keV 
luminosity (Figure \ref{pfreefig}). Again, we calculated the Kendall's $\tau$ coefficient 
to test for rank correlations between each parameter and  $L_{\rm X}$. The resulting Kendall's $\tau$ correlation coefficients were: 0.6 and 0.2, 
with $p$-values of: 0.02 
and 0.5 for $T_{\rm in}$ and $p$, respectively. These suggest that there may be a strong rank correlation between inner disk temperature and 
X-ray luminosity, although it is not highly significant (at the $3 \sigma$ level), and the weighted mean inner 
disk temperatures of both the BHBs and ULXs agree within the uncertainties ($\overline{T_{\rm in}} = 1.19 \pm 0.06$ and $1.26 \pm 0.05~{\rm keV}$ 
for the BHBs and ULXs, respectively). 
In order to produce such similar temperature spectra over an order of magnitude in luminosity, the normalization, hence 
the inner radius of the disk, 
varies between BHBs and ULXs. Given the uncertainties in inclinations and distances only approximate values of, 
or limits on, the inner disk radii can be calculated, however these generally suggest slightly larger black holes in the ULXs. 
Alternatively, this could be consistent with a scenario where the accretion disk becomes strongly advective, hence radiation trapping 
makes it unobservable, within a radius corresponding to a fixed temperature. 
Then, increases in $L_{\rm X}$ would correspond to increases in the characteristic inner disk radius, while the observed inner disk 
temperature remains roughly constant. 

Spectral shape is not the only diagnostic of an advective, slim disk. While standard disks evolve along a $L \propto T^4$ trend, 
slim disks are expected to follow a shallower slope (e.g., \citealt{watarai_etal_2001}). There is the suggestion that the ULXs 
may follow a shallower trend than the BHBs in the left-hand panel of Figure \ref{pfreefig}. However, this is not conclusive 
as there is no strong evidence that all of these BD ULXs have the same black hole masses or inclinations. A thorough study 
of this potential trend in individual sources is beyond the scope of this work.

\begin{table*}
\centering
\caption{A thin accretion disk around a Kerr black hole}
\label{kerrbb}
\begin{tabular}{ccccccccc}
\hline
Source & Obs ID & ${N_{\rm H}}$ & $a^{a}$ & $i^{b}$ & ${M_{\rm BH}}^{c}$ & ${\dot{M}_{\rm eff}}^{d}$ & $l^e$ & $\chi^2/{\rm dof}$ \\
 & & $(10^{22}~{\rm cm^{-2}})$ & & (deg.) & ({\Msun}) & $(10^{18}~{\rm g~s^{-1}})$ & & \\
\hline
\multicolumn{9}{c}{Observed BHBs}\\
GX~339$-$4 & {\xmm}-1   & 0.69 & 0.35   & 57     & 7.5    & 3.1 & 0.21 & (3656.0/323) \\
           & {\xmm}-2   & 0.68 & \ldots & \ldots & \ldots & 3.1 & 0.21 & \ldots \\
LMC X-3    & {\xmm}-1   & 0    & 0.80   & 48     & 7.0    & 3.4 & 0.41 & (20997.7/943) \\
           & {\swift}-1 & 0    & \ldots & \ldots & \ldots & 1.9 & 0.23 & \ldots \\
           & {\xmm}-2   & 0    & \ldots & \ldots & \ldots & 1.9 & 0.23 & \ldots \\
LMC X-1    & {\xmm}-1   & 0.61 & 0.64   & 56     & 10.9   & 2.2 & 0.14 & (3088.1/1235) \\
           & {\swift}-1 & 0.60 & \ldots & \ldots & \ldots & 2.7 & 0.16 & \ldots \\
           & {\swift}-2 & 0.61 & \ldots & \ldots & \ldots & 2.5 & 0.15 & \ldots \\
\hline

\multicolumn{9}{c}{Resampled BHBs}\\
GX~339$-$4 & {\xmm}-1   & $0.69 \pm 0.01$           & $0.4^{+0.1}_{-0.3}$    & $53^{+17}_{-7}$ & 7.5   & $3^{+2}_{-1}$           & 0.21 & 1158.4/1125 \\
           & {\xmm}-2   & $0.68 \pm 0.01$           & \ldots                 & \ldots          & \ldots & $3^{+2}_{-1}$          & 0.21 & \ldots \\
LMC X-3    & {\xmm}-1   & [0] & $0.80^{+0.08}_{-0.02}$ & $48 \pm 1$ & 7.0 & $3.3 \pm 0.1$ & 0.39 & (1714.3/1377) \\
           & {\swift}-1 & [0] & \ldots & \ldots & \ldots & $1.92 \pm 0.08$ & 0.23 & \ldots \\
           & {\xmm}-2   & [0] & \ldots & \ldots & \ldots & $1.83^{+0.08}_{-0.07}$ & 0.22 & \ldots \\
LMC X-1    & {\xmm}-1   & $0.57 \pm 0.01$ & $0.94^{+0.05}_{-0.50}$ & $38^{+31}_{-8}$ & 10.9  & $1.1^{+3.1}_{-0.3}$    & 0.13 & (1781.8/1344) \\
           & {\swift}-1 & $0.58 \pm 0.01$ & \ldots & \ldots & \ldots & $1.3^{+3.7}_{-0.3}$ & 0.15 & \ldots \\
           & {\swift}-2 & $0.61 \pm 0.01$ & \ldots & \ldots & \ldots & $1.3^{+3.5}_{-0.3}$ & 0.14 & \ldots \\
\hline

\multicolumn{9}{c}{ULXs}\\
M31 ULX2      & \ldots & $0.345^{+0.007}_{-0.013}$      & $0.990^{+0.009}_{-0.011}$ & $47^{+1}_{-2}$       & $46^{+6}_{-2}$       & $5.0^{+0.4}_{-1.0}$    & 0.20 & (420.6/307) \\
M31 ULX1      & \ldots & 0                              & $>0.9995$                 & $64 \pm 1$           & $80 \pm 4$           & $4.67^{+0.24}_{-0.02}$ & 0.15 & (240.6/136) \\
NGC 253 XMM2  & \ldots & $0.071 \pm 0.004$              & $>0.98$                   & $57^{+2}_{-6}$       & $42^{+2}_{-9}$       & $6.6^{+1.5}_{-0.4}$    & 0.33 & 357.3/324 \\
NGC 253 ULX2  & \ldots & $0.22 \pm 0.02$                & $0.91^{+0.06}_{-0.29}$    & $> 80$               & [50]                 & $40^{+50}_{-20}$       & 0.89 & 325.4/345 \\
M33 X-8       & \ldots & 0.026                          & 1.0                       & 53                   & 42                   & 6.6                    & 0.33 & (720.2/161) \\
NGC 2403 X-1  & \ldots & $0.204 \pm 0.006$              & $>0.97$                   & $57 \pm 4$           & $70^{+10}_{-20}$     & $7^{+4}_{-2}$          & 0.26 & 320.8/328 \\
NGC 4736 ULX1 & \ldots & [0]                            & $>0.989$                  & $57^{+1}_{-15}$      & $88^{+8}_{-36}$      & $7.5^{+0.8}_{-0.3}$    & 0.18 & (400.3/239) \\
\hline \\
\end{tabular}
\begin{minipage}{\linewidth}
As Table \ref{diskbb}, but for the absorbed {\sc kerrbb} model ($\textsc{tbabs} \times \textsc{kerrbb}$ in {\sc xspec}). For the BHBs 
the model was fitted to all observations simultaneously. The absorption column 
and accretion rate were allowed to vary between observations of a single source, but the black hole mass, spin, and inclination were fixed 
between epochs, and are shown for only the first observation of each source.
Notes:
$^a$black hole spin;
$^b$disk inclination;
$^c$black hole mass, for the BHBs this was fixed to the value, or the value in the middle of the range, given in 
Table \ref{BHB_catalogue};
$^d$effective mass accretion rate;
$^e$approximate Eddington ratio calculated as $\eta  M_{\rm dd} c^2 / (M_{\rm BH} \times 1.3 \times 10^{38} \ergs~M_{\odot}^{-1})$, where $\eta$ 
is the standard radiative efficiency of a black hole, which was calculated for each source using the 
best fitting value of the spin parameter shown here 
(note that Eddington ratio is not a parameter of the model).
\end{minipage}
\end{table*}

\begin{turnpage}
\begin{table*}
\centering
\caption{A thin accretion disk around a Kerr black hole plus a power-law}
\label{simpl*kerrbb}
\begin{tabular}{cccccccccccc}
\hline
Source & Obs ID & ${N_{\rm H}}$ & $\Gamma^a$ & ${f_{\rm sc}}^b$ & $a$ & $i$ & ${M_{\rm BH}}$ & ${M_{\rm dd}}$ & $l$ & $\chi^2/{\rm dof}$ & ${\Delta \chi^2/\Delta {\rm dof}}^c$ \\
 & & $(10^{22}~{\rm cm^{-2}})$ & & & & (deg.) & ({\Msun}) & $(10^{18}~{\rm g~s^{-1}})$ & & \\
\hline
\multicolumn{12}{c}{Observed BHBs}\\
GX~339$-$4   & {\xmm}-1   & 0.69 & 1.1 & 0.016 & 0.38   & 53     & 7.5    & 2.9 & 0.20 & (1315.2/319)  & $-2340.8/4$ \\
             & {\xmm}-2   & 0.69 & 2.3 & 0.025 & \ldots & \ldots & \ldots & 2.8 & 0.19 & \ldots        & \ldots \\
LMC X-3      & {\xmm}-1   & 0    & 4.5 & 0.42  & 0.65   & 45     & 7.0    & 3.6 & 0.34 & (19085.0/937) & $-1912.8/6$ \\
             & {\swift}-1 & 0    & 4.5 & 0.57  & \ldots & \ldots & \ldots & 2.0 & 0.19 & \ldots        & \ldots \\
             & {\xmm}-2   & 0    & 4.5 & 0.39  & \ldots & \ldots & \ldots & 2.0 & 0.19 & \ldots        & \ldots \\
LMC X-1      & {\xmm}-1   & $0.648^{+0.002}_{-0.005}$ & $4.5 \pm 0.4$          & $0.12^{+0.04}_{-0.02}$       & $0.44^{+0.04}_{-0.08}$ & $58^{+3}_{-2}$ & 10.9   & $2.9^{+0.4}_{-0.2}$ & 0.14 & (2294.0/1229) & $-794.1/6$ \\
             & {\swift}-1 & $0.548^{+0.002}_{-0.003}$ & [1]                    & $0.4552^{+0.0128}_{-0.0004}$ & \ldots                 & \ldots         & \ldots & $6.0^{+1.1}_{-0.2}$ & 0.30 & \ldots        & \ldots \\
             & {\swift}-2 & $0.544 \pm 0.006$         & $4.48^{+0.21}_{-0.03}$ & $0.606^{+0.070}_{-0.006}$    & \ldots                 & \ldots         & \ldots & $2.6^{+0.4}_{-0.2}$ & 0.13 & \ldots        & \ldots \\             
\hline

\multicolumn{12}{c}{Resampled BHBs}\\
GX~339$-$4 & {\xmm}-1 & $0.692 \pm 0.007$ & [1]                 & $0.267^{+0.176}_{-0.005}$ & $0.35^{+0.02}_{-0.23}$ & $47.7 \pm 0.4$ & 7.5    & $3.67^{+4.48}_{-0.02}$ & 0.24 & 1142.1/1121 & $-16.3/4$ \\
           & {\xmm}-2 & $0.701 \pm 0.007$ & $3.7^{+0.2}_{-2.6}$ & $0.14 \pm 0.01$           & \ldots                 & \ldots         & \ldots & $2.58^{+1.41}_{-0.02}$ & 0.17 & \ldots      & \ldots \\

LMC X-3 & {\xmm}-1   & 0 & $3.52^{+0.08}_{-0.09}$ & $0.61^{+0.14}_{-0.02}$ & $0.367^{+0.007}_{-0.102}$ & $44.5 \pm 0.3$ &  7.0   & $3.87 \pm 0.03$        & 0.28 & (1644.0/1371) & $-70.3/6$ \\
        & {\swift}-1 & 0 & $4.16 \pm 0.07$        & [1]                    & \ldots                    & \ldots         & \ldots & $2.15^{+1.75}_{-0.02}$ & 0.16 & \ldots        & \ldots \\
        & {\xmm}-2   & 0 & $4.23^{+).09}_{-0.08}$ & $0.8^{+0.1}_{-0.2}$    & \ldots                    & \ldots         & \ldots & $2.14^{+0.32}_{-0.02}$ & 0.15 & \ldots        & \ldots \\
        
LMC X-1      & {\xmm}-1   & $0.633^{+0.007}_{-0.008}$ & $1.3^{+0.5}_{-0.3}$ & $0.013^{+0.002}_{-0.004}$ & $0.926^{+0.003}_{-0.002}$ & $31.1^{+9.7}_{-0.8}$ & 10.9      & $1.183^{+0.007}_{-0.016}$ & 0.13 & (1676.0/1338) & $-105.8/6$ \\
             & {\swift}-1 & $0.605^{+0.007}_{-0.008}$ & $4.5^{+0.5}_{-0.1}$ & $0.22^{+0.01}_{-0.02}$    & \ldots                    & \ldots               & \ldots    & $1.31^{+0.29}_{-0.01}$    & 0.14 & \ldots        & \ldots \\
             & {\swift}-2 & $0.560^{+0.007}_{-0.008}$ & $4.6^{+0.2}_{-0.1}$ & $0.56 \pm 0.02$           & \ldots                    & \ldots               & \ldots    & $1.076^{+0.018}_{-0.006}$ & 0.12 & \ldots        & \ldots \\
\hline

\multicolumn{12}{c}{ULXs}\\
M31 ULX2 & \ldots & $0.40 \pm 0.02$ & $2.6^{+0.7}_{-1.3}$ & $0.2^{+0.7}_{-0.1}$ & $>0.5$ & [42] & $>30$ & $5^{+66}_{-1}$ & 0.13 & 308.5/305 & $-112.1/2$ \\
M31 ULX1 & \ldots & [0] & [45] & [$3 \times 10^{-19}$] & $>0.9995$ & $64 \pm 1$ & $81^{+3}_{-4}$ & $4.66^{+0.24}_{-0.02}$ & 0.15 & (240.6/134) & $0.0/3$ \\
NGC 253 XMM2 & \ldots & $0.089^{+0.003}_{-0.006}$ & $3.24^{+0.26}_{-0.04}$ & $>0.5$ & $0.9949^{+0.0006}_{-0.1066}$ & $57.7^{+0.5}_{-12.3}$ & $101^{+69}_{-2}$ & $5^{+12}_{-1}$ & 0.09 & 325.9/322 & $-31.4/2$ \\
NGC 253 ULX2 & \ldots & $0.22 \pm 0.01$ & [1] & $<0.09$ & $0.91 \pm 0.06$ & $>80$ & $50^{+40}_{-20}$ & $40^{+40}_{-20}$ & 0.90 & 325.3/343 & $-0.1/2$ \\
M33 X-8 & \ldots & $0.042^{+0.002}_{-0.001}$ & $3.4^{+0.2}_{-0.4}$ & $>0.3$ & $>0.5$ & $57^{+2}_{-4}$ & $80 \pm 10$ & $5.3^{+1.4}_{-0.9}$ & 0.13 & (310.7/159) & $-409.5/2$ \\
NGC 2403 X-1 & \ldots & $0.207 \pm 0.004$ & $1.03^{+3.84}_{-0.03}$ & $0.049^{+0.025}_{-0.007}$ & $>0.98$ & $57.8^{+0.4}_{-5.0}$ & $78 \pm 1$ & $7.64^{+1.44}_{-0.05}$ & 0.25 & 319.8/326 & $-1.0/2$ \\
NGC 4736 ULX1 & \ldots & 0 & $3.23^{+0.14}_{-0.04}$ & $[1]$ & $0.9962^{+0.0004}_{-0.0323}$ & $53.2^{+0.7}_{-6.0}$ & $188^{+62}_{-5}$ & $4.7^{+18.5}_{-0.8}$ & 0.052 & (318.9/237) & $-81.4/2$ \\

\hline \\
\end{tabular}
\begin{minipage}{\linewidth}
As Table \ref{kerrbb}, but with an additional power-law component 
($\textsc{tbabs} \times \textsc{simpl} \ast \textsc{kerrbb}$ in {\sc xspec}).
The parameters of the {\sc simpl} power-law were allowed to vary between observations of a single source. 
Notes: 
$^a$power-law spectral index;
$^b$fraction of the intrinsic disk flux that is scattered into the power-law;
$^c$improvement in the fit statistic relative to the absorbed {\sc kerrbb} model alone.
\end{minipage}
\end{table*}
\end{turnpage}

\subsection{A thin disk around a Kerr black hole}

The final spectral model was of a thin accretion disk around a Kerr black hole ({\sc kerrbb} in {\sc xspec}). 
The free parameters of this model were: black hole spin, inclination, black hole mass (fixed for the BHBs), and the effective 
mass accretion rate. Additional parameters that were fixed were: distance from the observer; 
$\eta$, the torque parameter, which was set to 0; spectral hardening factor, which was set to 
1.7; a flag for self-irradiation, which was fixed at 1 (i.e., included in the model); a flag for limb darkening, which was set to 0 (i.e., 
excluded from the model).
Theoretically, 
this model is appropriate for accretion rates up to $\sim 0.3$ Eddington, above which radiation pressure begins to inflate the disk 
(e.g., \citealt{mcclintock_etal_2006}) such that so-called slim disk models with radial advection are more appropriate. 
In the {\sc kerrbb} thin disk model, no single parameter affects 
normalization alone: increasing the spin or accretion rate results in a harder spectrum and higher bolometric luminosity 
while increasing the black hole mass softens the spectrum. 
Alternative General Relativistic accretion disk models exist (see e.g., \citealt{middleton_BHspin} for a summary) including {\sc bhspec} 
\citep{davis_etal_2005}. 
Unlike {\sc kerrbb}, which assumes a constant spectral hardening factor, {\sc bhspec} 
includes bound-free and free-free  opacities of abundant ion species, 
which can produce a slightly broader spectrum \citep{kubota_etal_2010}. 
However, \cite{hui_and_krolik_2008} report that there is little difference 
between the fits or resulting parameters when fitting {\sc kerrbb} and {\sc bhspec} to ULXs with disk-like spectra, so we only consider the {\sc kerrbb} 
model here. 

The fit statistics and parameters from the {\sc kerrbb} model are given in Table \ref{kerrbb}. For the BHBs, where we are analyzing 
multiple observations of each source, each of the spectra were fitted simultaneously with spin and inclination fixed between them. 
The model can be rejected for all of the real BHB data sets, and it is only acceptable in 1/3 of the resampled BHBs and 3/7 of the ULXs. 
When fitting the BHBs, we fixed the black hole mass to the values given in Table \ref{BHB_catalogue}. 
As an additional test, we refitted both the real and resampled spectra from GX~339$-$4 while allowing the black hole mass to 
vary within the quoted uncertainty range. 
For the real data, the improvement in the fit statistic was marginally significant based on an $f$-test (at the $2 \sigma$ level, but not the $3 \sigma$ level), although 
 the model could still be strongly rejected ($\chi^2/{\rm dof} = 3564.4/322$). Despite the slight improvement in fit, allowing the black 
hole mass to vary did not result in large changes ($\lesssim 10$\%) to the other model parameters. No such improvement in fit 
was found for the resampled data ($\Delta \chi^2 = -0.9$ for 1 fewer degree of freedom), and the model parameters 
were consistent within the $1 \sigma$ uncertainty regions with those from the fixed-mass model.

Although some of the model parameters appear to be very precise, we would caution against assuming them to be 
accurate, especially given the degeneracies between model parameters and the questionable physical basis of using a thin disk model. 
Additionally, when using spectral continuum fitting to determine meaningful estimates of the black hole spin, it is necessary to fix the inclination, mass, and distance 
to known values (e.g., \citealt{mcclintock_etal_2014}). We do not fix the inclination here, as 
it is not currently feasible in the ULXs, and measuring spin is not the purpose of this paper. 
Despite this, we note that the model results in spin and inclination estimates for GX~339$-$4 and 
LMC X-1 (from the resampled spectra) that agree with at least some reported results (although these are not always in consensus with each other, or 
particularly well constrained) while for 
LMC X-3, our values disagree with the literature: 
GX~339$-$4 - $a \approx 0.93$ \citep{miller_etal_2008},
$a > 0.97$ \citep{ludlam_etal_2015}, $a < 0.9$ \citep{kolehmainen_and_done_2010}, $i = 20$--$30\deg$
\citep{miller_etal_2004,reis_etal_2008,plant_etal_2014}, $i > 45 \deg$ 
\citep{kolehmainen_and_done_2010}; LMC X-3 - $a\approx 0.21$ \citep{steiner_etal_2014}, $i \approx 70 \deg$ \citep{orosz_etal_2014}; 
LMC X-1 - $a \approx 0.92$ \citep{gou_etal_2009}, $i \approx 36\deg$ \citep{orosz_etal_2009}. 
The disagreement in the case of LMC X-3 is hardly surprising, given that the spectral model was strongly rejected. 
While the range of parameters for the previous models did not vary significantly between the TD BHBs and BD ULXs, except 
for luminosity, this is not the case here. The fitted black hole masses in the ULXs varied between $\sim 40$--$80~M_{\odot}$, 
while values from Table \ref{BHB_catalogue} were used for the BHBs. Additionally, the black hole spins 
were relatively poorly constrained in the BHBs, but close to maximal in the ULXs, and mass accretion rates were significantly 
higher in the ULXs too. 
Despite the differences in parameters, the approximate Eddington ratios of the BHBs and ULXs largely overlapped with both classes 
typically falling between $\sim 0.1$--0.4. The sole exception to this was NGC 253 ULX2, which had a higher Eddington ratio of 
$\sim 0.9$, but we note that the black hole mass parameter could not be constrained by the data, so this estimate is subject 
to extremely large uncertainties.

Overall, the {\sc kerrbb} parameters suggested that the ULXs contained slightly larger black hole primaries than the BHBs, and they 
were at close to maximal spin. However, the {\sc kerrbb} model is rejected in 4 out of the 7 ULXs, so the variables of the model 
cannot necessarily be simply interpreted as the physical parameters of the ULX binary systems. Indeed, in the case of M31 ULX2, 
the black hole mass is clearly in disagreement with the constraints from luminosity limits based on a sub-Eddington spectral state 
when it is faint ($< 17~M_{\odot}$, \citealt{middleton_etal_2013}). 
Fixing the black hole to $10~M_{\odot}$ for this source 
resulted in a significantly worse fit statistic ($\chi^2/{\rm dof} = 469.9/308$ or $\Delta \chi^2 = 49.3$ for 1 additional degree of freedom) 
compared to the model with the black hole 
mass free to vary (which had a black hole mass of $46^{+6}_{-2}~M_{\odot}$, where the uncertainties are reported at the $1 \sigma$ level), 
with a corresponding 
$f$-test probability of $5.6 \times 10^{-9}$. 
Additionally, fixing the black hole mass resulted in the model having retrograde spin 
($a = -0.8 \pm 0.2$; cf. \citealt{middleton_etal_2014}) and 
a significantly higher mass accretion rate ($M_{\rm dd} = 28^{+6}_{-5} \times 10^{18}~{\rm g~s^{-1}}$, or $l \sim 2$). 
This may not be surprising since black hole accretion disks are not expected to remain geometrically thin as $l \rightarrow 1$.

\begin{figure*}
\begin{center}
\includegraphics[width=16cm]{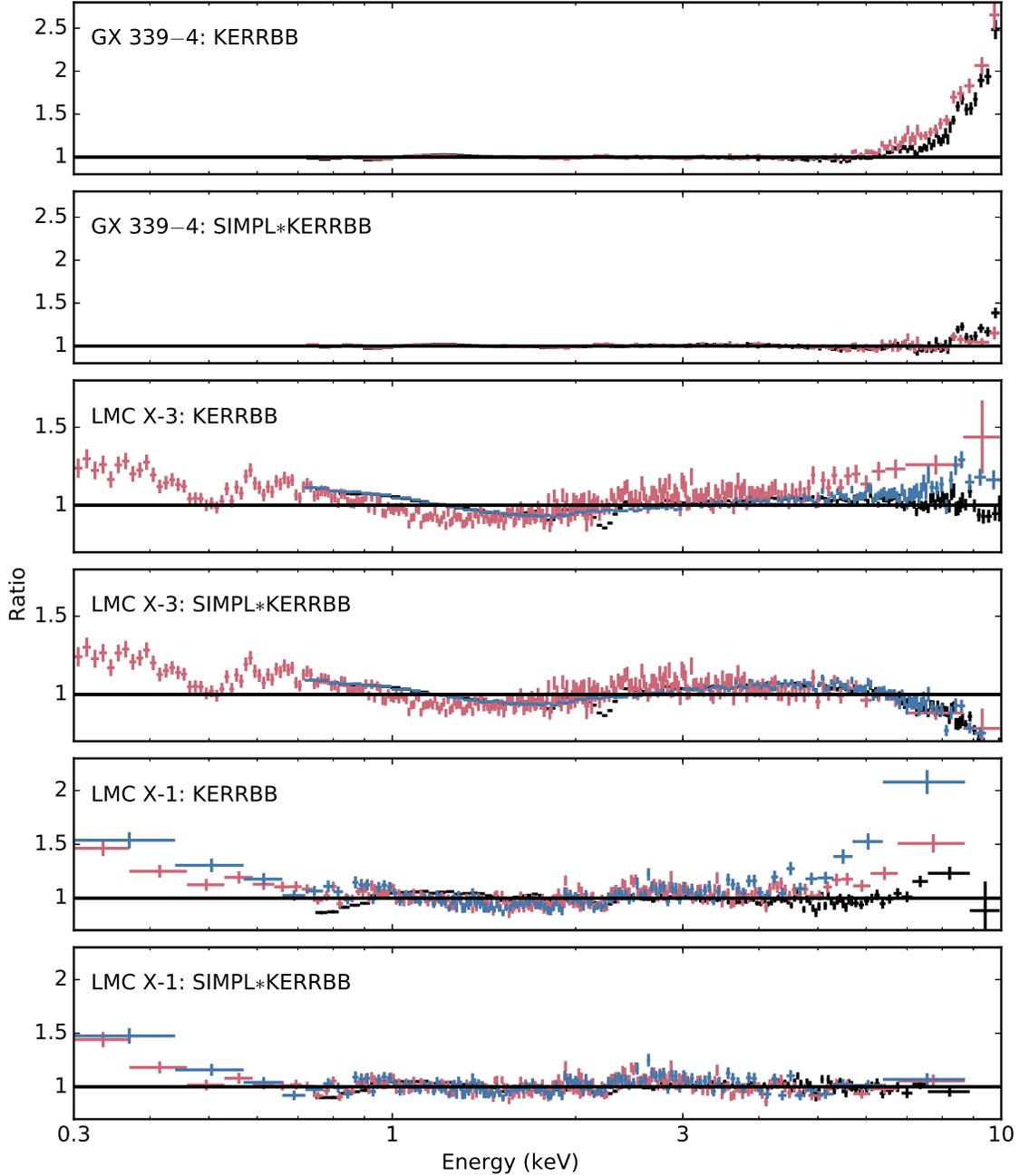}
\caption{Residuals to the best-fitting absorbed {\sc kerrbb} and {\sc simpl}{$\ast$}{\sc kerrbb} models for the real BHB data plotted as ratios. 
Colors correspond to the different observations, with black being the first, red the second, and blue the third observation in 
order of observation date as given in Table \ref{BHB_obs}. The data have been rebinned to $20 \sigma$ significance for clarity. 
It was evident that the absorbed {\sc kerrbb} model alone under-predicted the 5--10 keV flux in most of the BHB spectra, which may indicate 
the presence 
of a power-law-like tail, even in the {\xmmn} and {\swift} energy band. For GX~339$-$4 and LMC X-1 the addition of a {\sc simpl} power-law component 
reduced the hard residuals. The model parameters indicated that GX~339$-$4 and LMC X-1 {\xmm}-1 were in the TD state, while 
LMC X-1 {\swift}-1 and -2 may have steep power-law spectra. However, for LMC X-3 the combination of a thin disk and {\sc simpl} power-law 
spectrum resulted in large negative residuals at high energies, bringing the validity of the model into question.}
\label{BHB_residuals}
\end{center}
\end{figure*}

\subsubsection{Additional power-law emission}\label{spl}

Although we attempted to exclude spectra with obvious high energy power-law emission, it was evident from the spectral residuals from the {\sc kerrbb} 
model fitted to the real BHB spectra that some may deviate from pure disk spectra (Figure \ref{BHB_residuals}). 
However, in the BHB with the largest 
deviation between the model and data, GX~339$-$4, the {\sc kerrbb} model was able to adequately fit the $\sim$~25000 count resampled spectra. 
This indicates that relatively weak power-law emission (as in the TD state) may not be detected if present in similar quality observations of nearby ULXs. 
Regardless of this, we added a power-law 
component to the {\sc kerrbb} model using the {\sc simpl} convolution model in {\sc xspec} and tested whether this improved the fits to the real 
BHB spectra (Table \ref{simpl*kerrbb}). 
The {\sc simpl} model has parameters of spectral index; the fraction of the input emission that is reprocessed into the power-law; and a flag for 
up-scattering only, or both up- and down-scattering, which was set to up-scattering only.
Indeed, adding this component resulted in large improvements in the fit statistic: $\Delta \chi^2 = -2340.8$, $-1912.8$, and $-794.1$, 
for 4, 6, and 6 fewer degrees of 
freedom, in GX~339$-$4, 
LMC X-3, and LMC X-1 respectively, although all of these fits were still rejected at $3 \sigma$ significance.

As with the previous model, we used the real and resampled spectra from GX~339$-$4 to test whether allowing the black hole mass to 
vary within 
the range given in Table \ref{BHB_catalogue} significantly altered the fits. 
No significant improvements in the fit statistics were found (at the $>95$\% level, based on an $f$-test), and the model parameters did 
not vary significantly from the 
fixed-mass model (they were within the $1 \sigma$ uncertainty regions where calculated, or varied by $<10$\%).

For both observations of GX~339$-$4 only $\sim 2$\% of the model intrinsic disk flux was emitted as a power-law, which had 
a spectral index of $\sim$~1--2 
(although a higher scattering fraction and softer spectral index were found for the sub-sampled data). 
These values are consistent with fits to sources in the TD state (cf. \citealt{steiner_etal_2009}). 
We generally found extremely soft power-law spectral indices in LMC X-3 and X-1, with large fractions of the input intrinsic disk 
flux being scattered 
($\Gamma \sim 4.5$ and scattered fraction $f_{\rm sc} \sim 0.3$--0.6; LMC X-1 {\it Swift}-1 had a very hard power-law resulting 
in a large fraction of the intrinsic model disk flux being scattered out of the 0.3--10 keV band, but 
this was not well constrained), 
and as such we caution against interpreting these physically. 
Never-the-less, if we do attempt to interpret them physically, they may suggest that LMC X-3, and LMC X-1 during the {\swift} observations, were in 
a steep power-law (cf. \citealt{steiner_etal_2009}) or possibly hyper-soft \citep{uttley_and_klein-wolt_2015} state. 
However, an inspection of the 
hard residuals to LMC X-3 indicated that the {\sc simpl} power-law was not a good fit to the data (Figure \ref{BHB_residuals}), 
so it may not be possible 
to trivially interpret this as a physical component. Additionally, 
LMC X-3 has previously nearly always been observed with a TD spectrum (e.g., \citealt{steiner_etal_2010}). As such, we do not rule out 
a TD spectrum in this source. On the other hand, LMC X-1 has previously been seen with both TD and steep power-law spectra 
(e.g., \citealt{alam_etal_2014}) and, for the {\swift} observations, the {\sc simpl} power-law eliminated the residuals above $\sim 5$~keV. 
It is possible that LMC X-1 was in the steep power-law state during the {\swift} observations, thus not dominated by emission from 
the accretion disk in the 0.3--10 keV energy range. When we eliminated these observations and only fitted LMC X-1 {\xmm}-1, the {\sc kerrbb} 
model was marginally acceptable at the $3 \sigma$ level for the resampled data ($\chi^2/{\rm dof} = 651.6/555$, corresponding to a null 
hypothesis likelihood of 0.997), although broad disk and two-component models had fit statistics that were lower still.

We also tested whether modifying the ULX {\sc kerrbb} fits with the {\sc simpl} convolution model could improve the fits (Table \ref{simpl*kerrbb}). 
This additional component resulted in improvements in the fit statistic of $\Delta \chi^2 = -112.1$, 0, $-409.5$, and $-81.4$ for 2 
fewer degrees of freedom, for M31 ULX2, M31 ULX1, M33 X-8, and NGC 4736 ULX1 respectively, where the {\sc kerrbb} alone was rejected. 
Notably, this model resulted in a statistically 
acceptable fit for M31 ULX2 ($\chi^2/{\rm dof} = 308.5/305$), where the parameters indicated a 
$\sim 70$~{\Msun} black hole in the TD state with only $\sim 0.7$\% of the intrinsic disk flux being reprocessed 
in to a hard ($\Gamma \sim 1.4$) power-law. 
Also, the {\sc simpl} model parameters indicated that M33 X-8 and NGC 4736 ULX1 may have spectra similar to steep power-law 
state BHBs ($\Gamma \approx 2.9$, $f_{\rm sc} \approx 0.3$ and $\Gamma \approx 3.1$, $f_{\rm sc} \approx 0.6$ respectively), 
however these fits were still rejected at $> 3 \sigma$ significance ($\chi^2/{\rm dof} = 312.7/159$ and 326.1/237) 
indicating that the BD spectra in these ULXs is unlikely to be due to an unresolved hard power-law component.

\begin{figure}
\begin{center}
 \includegraphics[width=7cm]{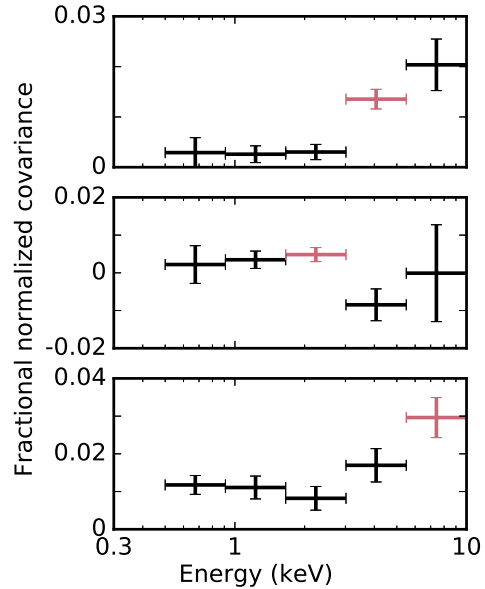}
\caption{Fractional normalized covariance spectra of (top) GX~339$-$4 {\xmm}-1, (center) {\xmm}-2, and (bottom) LMC X-3 {\xmm}-1. Red data points correspond to 
reference bands, where fractional rms variability is plotted instead. Constant fracitonal normalized covariance 
can be rejected at $3 \sigma$ significance in GX~339$-$4 {\xmm}-1, indicating the presence of a second spectral 
component.}
\label{bhb_covar}
\end{center}
\end{figure}

\begin{figure}
\begin{center}
 \includegraphics[width=7cm]{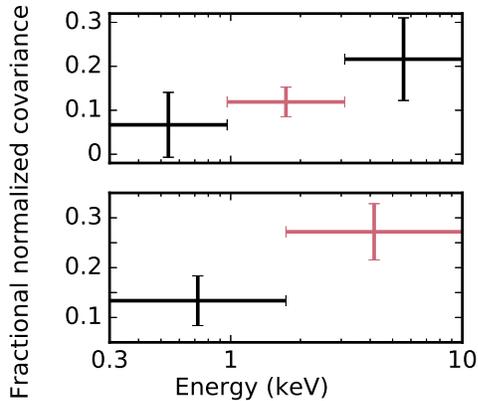}
\caption{Fractional normalized covariance spectra of (top)  NGC 253 XMM2 and (bottom) NGC 4736 ULX1. Red data points correspond to 
reference bands, where fractional rms variability is plotted instead. Both of the sources were consistent (at  $3 \sigma$ significance) with having 
constant fractional normalized covariance, thus there is no strong requirement for two-component 
models.}
\label{covar}
\end{center}
\end{figure}

\subsection{Additional emission components in M33 X-8}

While  most of the ULXs could be well fitted by 
at least some of the spectral models, no statistically acceptable model was found for M33 X-8. This could potentially indicate additional spectral
emission components were present in this source. Indeed, \cite{laparola_etal_2003} reported a broad Gaussian emission line at 
$\sim 0.9$ keV and $\sim 0.2$ keV thermal emission from a hot plasma in the {\chandra} ACIS spectrum of M33 X-8, and \cite{owen_and_warwick_2010} 
reported that extended emission in the galaxy could be fitted by a dual thermal plasma spectrum, with $kT \sim 0.2$ and 0.6 keV. The addition 
of a broad 
Gaussian emission line resulted in the fit statistic being improved by $\Delta \chi^2 = -115.2$, $-58.4$, $-80.7$, and $-308.1$ for 
3 fewer degrees of freedom, for the MCD plus power-law, MCD plus Comptonized emission, $p$-free disk, and {\sc kerrbb} models respectively. 
However, the parameters of the emission line  were dependent on the continuum model. For the first 3 models the line energy and width 
were $E \sim 1.1~{\rm keV}$ and $\sigma \sim 0.2~{\rm keV}$, while for the {\sc kerrbb} model they were $E \sim 0.8~{\rm keV}$ 
and $\sigma \sim 0.5~{\rm keV}$. An additional $\sim 0.2~{\rm keV}$ mekal plasma component did not further improve any of the fits.

\subsection{X-ray timing}

We have been fitting the BHB and ULX spectra with physical and phenomenological models of emission from accretion disks. 
As a further test of whether single component spectral models were appropriate for the ULXs, we examined their timing properties 
by extracting fractional normalized covariance spectra. 
We were able to constrain the fractional variability in one energy bin, thus could extract covariance, for two of the ULXs 
(NGC 253 XMM2 and NGC 4736 ULX1) and three of the BHB observations (GX~339$-$4 {\xmm}-1, -2, and LMC X-3 {\xmm}-1).
The resulting fractional normalized covariance spectra are shown in Figure \ref{bhb_covar} and \ref{covar}. 
Constant fractional normalized covariance could not be strongly ruled out for any of the ULXs (
$\chi^2 / {\rm dof}  = 1.6/2$ and 3.4/1
for NGC 253 XMM2 and NGC 4736 ULX1, respectively, when fitted with a constant), 
albeit with the caveat that covariance could only be extracted from two of the ULXs, and even for these several values were admittedly 
poorly constrained.
We note that \cite{middleton_etal_2012} reported fractional covariance in M31 ULX1 that varied with energy, but for different energy and 
temporal binning schemes than we used here. 
For the BHBs, a constant value of fractional normalized covariance could be rejected at $3 \sigma$ significance in a 
single observation: {\xmm-1} of GX~339$-$4 
($\chi^2/{\rm dof} = 31.8/4$). Constant values were not rejected at $3 \sigma$ significance 
in GX~339$-$4 {\xmm}-2 or LMC X-3 {\xmm}-1 ($\chi^2/{\rm dof} = 9.5/4$ and 13.5/4 respectively).

The covariance spectra from observation {\xmm}-1 of GX~339$-$4 peaked in the hard band as would be expected 
if a hard, variable power-law was contributing to the energy spectrum. 
We tested whether the covariance spectrum from this observation was consistent with the shape of the hard power-law in the 
{\sc simpl}$\ast${\sc kerrbb} model. 
We converted the (non-fractional) normalized covariance spectra to {\sc xspec} spectra using {\sc flx2xsp} and 
fitted it with the {\sc simpl} power-law.\footnote{To isolate the power-law from the convolution model we added an additional 
{\sc kerrbb} component, i.e., $\textsc{tbabs} \times \textsc{tbabs} \times (\textsc{simpl} \ast \textsc{kerrbb} + \textsc{kerrbb})$, 
with all parameters except normalization equal in both of the accretion disk components. The normalization of the second {\sc kerrbb} 
component was set equal to $f_{\rm sc}-1$.} An additional constant was added to the model to allow the normalization to vary, 
while all other model parameters were fixed to the best fitting values from the spectral fitting. The resulting fit statistic was 
$\chi^2/{\rm dof} = 7.9/4$, which cannot 
be rejected even at $2 \sigma$ significance. 

\section{discussion}\label{discussion}

\begin{figure*}
\begin{center}
\includegraphics[width=12cm]{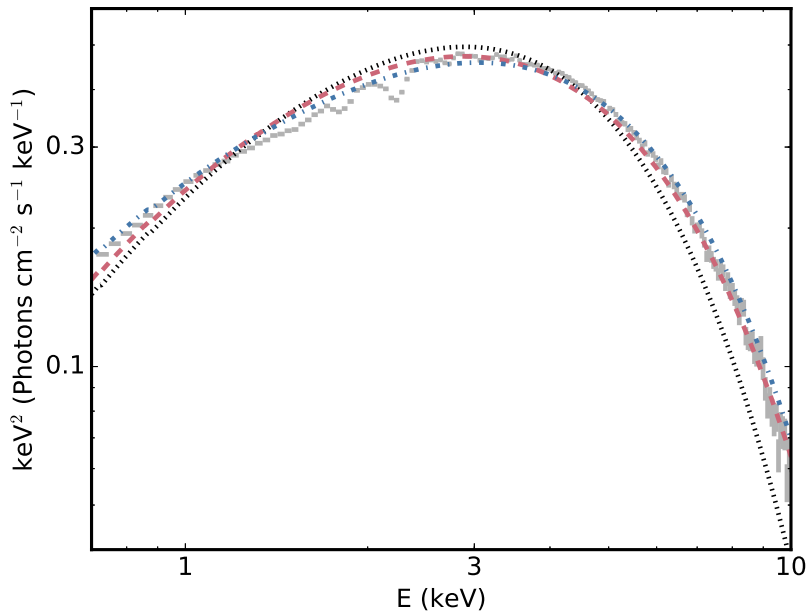}
\caption{The spectrum from LMC X-3 {\xmm}-1, which is shown as an example of a TD BHB spectum that is broader than 
expected for a thin disk. 
The spectrum is over-plotted with the three disk models used in this paper. These are (in order 
of increasing breadth): an MCD (black dotted line), 
{\sc kerrbb} (red dashed line), and a $p$-free disk (blue dot-dashed line). Clearly the MCD and {\sc kerrbb} spectral models 
are insufficiently broad, and a broader spectrum, such as the $p$-free disk, improves the fit.}
\label{spec_eg1}
\end{center}
\end{figure*}

We have fitted the energy spectra 
of sub-Eddington ($l \gtrsim 0.1$) BHBs in the 0.3$-$10~keV energy range with accretion disk emission models.
The models were for the most part insufficient to reproduce the spectra from the BHBs unless they were sub-sampled to be comparable to the ULX sample.
However, it was clear that many of the BHB spectra, even at these low Eddington ratios, were broader than expected 
from standard geometrically-thin optically-thick disks, i.e., from a purely thermal MCD disk model. 
This is evident in the example of LMC X-3 {\xmm}-1 (Figure \ref{spec_eg1}) and in the spectral residuals when LMC X-3 and X-1 
are fitted with the {\sc kerrbb} model (Figure \ref{BHB_residuals}). 
In GX~339$-$4 some of this broadening is probably due to a hard power-law component, 
however there is no strong evidence of this in any other object.
In fact, the fit statistic improved significantly in other sources when a power law or Comptonization component was added to 
the spectral model, 
but we interpret this strictly phenomenologically as a means of broadening the model spectrum rather than as a true second spectral component.
A similar conclusion was reached recently by \citet{kolehmainen_etal_2011} for GX~339$-$4.

More importantly, the spectral shape of our sample of BD ULXs could not be significantly distinguished from the BHBs in this study. 
This is rather interesting as 
it potentially weakens arguments for the BD and TD sources being in 
physically distinct states. 
Theoretically, the standard thin disk is expected to transition to a 
slim accretion disk state at high luminosities where radiation pressure causes the scale height of the disk to increase 
and advection of radiation to alter the emissivity structure of the disk. 
Such advection-dominated disks have spectra that are broader and less peaked 
than standard geometrically thin disks 
(for more detailed discussions of slim disks see e.g., 
\citealt{abramowicz_etal_1988,sadowski_2009}). 
Indeed, the $p$-free disk approximation of an advection-dominated disk was sufficient to reproduce most of the BD ULX and the resampled BHB spectra. 
However, accretion disks are only thought to become sufficiently radiation pressure dominated to 
enter the slim disk regime above $l \sim 0.3$ \citep{mcclintock_etal_2006,steiner_etal_2010,straub_etal_2011} 
whereas the BHB spectra appear to be broadened at values as low as $l \sim 0.1$. 
We note that spectral shape is not the only difference between thin and slim disks, and a further test of the nature of these accretion 
disks could come from the luminosity -- temperature relation. Standard thin accretion disks follow 
$L \propto T^4$, while the luminosity of a slim disk would increase less steeply with temperature ($L \propto T^2$; 
\citealt{kubota_and_makishima_2004}). 
However, to avoid degeneracies with black hole masses and inclinations, this is best studied in individual objects, so is beyond the scope of 
this work.

As BD-like spectra are seen at low Eddington ratios, such spectra cannot be considered as a unique signature of $\sim$ Eddington rate 
accretion as we had anticipated. Therefore, it is not clear how to estimate the innermost stable circular orbit (and hence constrain the compact object mass) 
from the normalization of the disk models, as different coefficients are often 
used for standard \citep{shimura_and_takahara_1995, kubota_etal_1998, davis_etal_2005} and slim regimes \citep{vierdayanti_etal_2008}, 
perhaps with some continuous range in between. 
Despite this, given the similar spectra of the BHB and BD ULX sources, it is tempting to speculate whether they may all 
be at similar Eddington ratios, $l \sim 0.1$--0.4. 
Theoretically, the most physically representative model of this state applied in this work is the {\sc kerrbb} model.
Although this model is statistically ruled out for four of the seven ULX observations, this is also the case in two of the three 
sub-sampled BHBs, which have known Eddington ratios of $l \approx 0.1$--0.4. 
The {\sc kerrbb} model parameters suggest that the BD ULXs contain unusually massive black hole primaries ($\sim 40$--90 \Msun) 
with close to maximal spin (cf. \citealt{hui_and_krolik_2008}). 
To put this in to context, the recent first gravitational wave detection was from the merger of two black holes in
a binary with similarly large masses of $\sim 29$  and 36 {\Msun}, but with lower initial spin ($a<0.7$) at least 
for the higher mass primary \citep{GW_detection}. 
It is possible that the intrinsic disk spectra that we study here are inherently broader than the models that we use, such that the 
{\sc kerrbb} fits overestimate spin, as this gives larger relativistic broadening.

There are several problems with strict acceptance of the parameter values (besides the poor fit statistics) that we obtain for the ULXs from the {\sc kerrbb} model. 
First, the black hole masses implied for many of these ULXs fall in to the `second mass gap' predicted for compact object evolution. This mass gap 
covers the range of 
$\sim$ 60--130 {\Msun} over which it is thought that stellar black holes do not form due to disruption by pair instability supernovae 
\citep{Marchant_etal_2016}. Furthermore, black holes below the mass gap may be limited to $<50$ {\Msun} by severe mass loss in pulsation pair 
instability supernovae (Belczynski et al. in prep.).
Also, the collapse of massive stars is only thought to produce black hole spins of $a < 0.75$--0.9 \citep{gammie_etal_2004}. 
Stellar remnant black holes undergoing Roche lobe overflow could be spun up \citep{podsiadlowski_etal_2003,fragos_and_mcclintock_2015}, but 
to achieve maximal spin an initially non-spinning black hole must have accreted $\gtrsim 75$\% of its current mass \citep{bardeen_1970,king_and_kolb_1999}. 
For a present day 50~{\Msun} black hole to accrete $>$37~{\Msun} would take $\sim 5 \times 10^8~{\rm years}$ at a constant rate of $5 \times 10^{18}~{\rm g~s^{-1}}$.
This is much greater than the $\lesssim 10^7$~year lifespan of the required donor star. 
Finally, if the BD ULXs contain systematically larger black holes than those seen in Galactic binaries, yet all 
are accreting at sub-Eddington rates, then 
it would be unlikely that BD ULXs could almost exclusively bridge the gap in X-ray luminosity between the BHBs at $L_{\rm X} < 10^{39} \ergs$ 
and the two-component ULXs  
typically seen at $L_{\rm X} > 3 \times 10^{39} \ergs$ \citep{sutton_etal_2013b}. If the BD ULXs are sub-Eddington and the two-component 
ULXs are at $l > 1$, then 
there is a missing population of sources at $l \sim 1$.

We note that the combination of high accretor mass and high spin in the {\sc kerrbb} model is necessary to maintain a high luminosity at 
the required temperature, but 
not to produce the spectral curvature. 
Using M31 ULX1 as an example, holding the black hole mass fixed at the fitted value and allowing the spin to assume values of $a = 0.9$, 0.5, and 0 
results in the 0.3$-$10 keV 
unabsorbed luminosity being reduced to $\sim 6$, 2, and $1 \times 10^{38} \ergs$ respectively.
On the other hand, 
the lack of lower spin BHs in the ULX sample could feasibly be a selection effect, since lower spin objects would also be softer and 
less efficient, hence require higher (and presumably rarer) mass accretion rates to be classified as ULXs. 

Since it is arguably unlikely that all of the BD ULXs contain unusually large stellar remnant black holes with extremely high spins, 
we instead consider the possibility that BD ULXs are a high Eddington 
ratio extension of the standard stellar-mass X-ray binaries exemplified by the sample of BHBs.
Then, why are the spectra of these two populations so similar?
Recently, 
\cite{straub_etal_2013} showed that the spectral shape of M31 ULX1 remained nearly unchanged 
over a range of luminosities from $l \sim 0.3$ to $\sim 0.9$ and, moreover, differed only slightly from the standard thin-disk model. 
We also find no clear evidence that the BD ULXs are significantly broader than the BHB spectra, so the BD ULXs could be close to $l \sim 1$ 
while the BHBs are nearer $l \sim 0.1 - 0.4 $. 
This suggests that there may be some important physics missing from the disk models at low $l$ that accounts for the broadening. 
That is,
the progression between the TD BHBs and BD ULXs cannot 
simply be explained by an initially thin accretion disk becoming geometrically slim and advection-dominated. Instead, at face-value, these 
results suggest that the reported 
BHBs somehow conspire to remain slim down to low $l$. 

Recent radiation magnetohydrodynamical simulations show that a disk at $l \sim 0.8$ can be thermally 
stabilized by magnetic pressure for an initial (net vertical flux) quadrupole field \citep{sadowski_2016}. 
The magnetic pressure is larger than the thermal (dominated by radiation) pressure, so the disk has a somewhat larger 
scale height than predicted from the Shakura-Sunyaev model. This will lower the density of the disk, which reduces the opacity, 
and gives a larger color temperature correction to the emitted spectrum. There are some stringent observational constraints on the 
average color temperature correction and its constancy with luminosity \citep{done_and_davis_2008}, but it may be possible 
to produce the observed subtle disk broadening. 
Alternatively, magnetically driven winds may quench energy from the disk \citep{miller_etal_2006}.
Still, in either scenario radiation pressure would have to smoothly 
take over from magnetic pressure/winds to produce the similar observed broadened accretion disk spectra in both low-$l$ BHBs and $l \sim 1$ BD ULXs.

\section{conclusions}

We have presented an X-ray spectral and timing study comparing samples of BD ULXs, which are thought to be accreting at around the Eddington rate,  
with relatively luminous but sub-Eddington TD BHBs. At least some of these TD BHBs exhibit BD-like spectra in the 0.3--10 
keV band but at order of magnitude lower luminosities than the ULXs. 
While it is tempting to speculate that these BHBs and ULXs are accreting at similar Eddington ratios, this would imply near-maximally-spinning 
massive stellar black holes in the BD ULXs, which cannot be trivially produced. 
Instead, if the BD ULXs are at $l \sim 1$, then the spectral broadening can feasibly be explained by advection in radiation pressure dominated, 
geometrically slim disks. However, this cannot be the case in the TD BHBs as the Eddington ratios are constrained to moderate values. Instead, 
this implies that some other physical mechanism, such as magnetic pressure, takes over from radiation pressure to produce broad spectra at 
lower Eddington ratios.

\acknowledgments

The authors thank the anonymous referee for useful comments. 
ADS acknowledges funding through a NASA Postdoctoral 
Program appointment at Marshall Space Flight Center, administered by 
Universities Space Research Association on behalf of NASA. 
TPR acknowledges funding from STFC as part of the consolidated 
grant ST/L00075X/1.

All data used in this paper are publicly available in NASA's High 
Energy Astrophysics Archive Research Center (HEASARC) archive, 
which is a service of the Astrophysics Science Division at NASA/GSFC 
and the High Energy Astrophysics Division of the Smithsonian Astrophysical Observatory.
This paper is mainly based on observations obtained with {\it Swift} and {\it XMM-Newton}, an 
ESA science mission with instruments and contributions directly 
funded by ESA Member States and NASA. 
It also makes use of {\it RXTE} ASM light curves, provided by 
the ASM/{\it RXTE} teams at MIT and at the RXTE SOF and GOF at NASA's GSFC.

{\it Facilities}: \facility{Swift}, \facility{XMM}, \facility{RXTE}

\bibliography{refs}
\bibliographystyle{apj}
\end{document}